\documentclass[showpacs,showkeys,twocolumn,superscriptaddress]{revtex4-2}
\usepackage{bm,graphicx,subfigure,amsmath}
\usepackage{longtable,multirow,color}

\begin{document}
\title{Novel phases in rotating Bose-condensed gas: vortices and quantum correlation}
\author{Mohd. Imran}\email{alimran5ab@gmail.com}
\affiliation{Department of Physics, Jamia Millia Islamia (A Central University), New Delhi 110025, India} 
\author{M. A. H. Ahsan}\email{mahsan@jmi.ac.in}
\affiliation{Department of Physics, Jamia Millia Islamia (A Central University), New Delhi 110025, India}
\begin{abstract}
We present the exact diagonalization study of rotating Bose-condensed gas interacting via finite-range Gaussian potential confined in a quasi-2D harmonic trap. The system of many-body Hamiltonian matrix is diagonalized in given subspaces of quantized total angular momentum to obtain the lowest-energy eigenstate employing the beyond lowest-Landau-level approximation. In the co-rotating frame, the quantum mechanical stability of angular momentum states is discussed for the existence of phase transition between the stable states of interacting system. Thereby analyzing the von Neumann entanglement entropy and degree of condensation provide the information about quantum phase correlation in the many-body states. Calculating the conditional probability distribution, we further probe the internal structure of quantum mechanically stable and unstable states. Much emphasis is put on finding the spatial correlation of bosonic atoms in the rotating system for the formation and entry of singly quantized vortices, and then organizing into canonical polygons with and without a central vortex at the trap center. Results are summarized in the form of a movie depicting the vortex patterns having discrete p-fold rotational symmetry with $p = 2,3,4,5,6$.
\end{abstract}
\maketitle
\section{INTRODUCTION}
\label{intro}  
Ever since the Bose-Einstein condensate (BEC) was realized in dilute vapors of ultra-cold (nano-kelvin) alkali Bose atoms \cite{aem95,dma95,bst95}, the response of these systems to rotation has attracted a lot of experimental and theoretical attention \cite{dgp99,ajl01,fs01,bpa10}.
With experimental versatility such as controllable density, effective dimensionality and tunable inter-particle interactions \cite{ias98,cgj10}; these dilute and inhomogeneous Bose condensates has become an extremely convenient system to investigate the characteristics of macroscopic quantum phenomena.
Subsequently intense studies have been initiated on the various aspects of rotating condensate \cite{ps02,jfa03}.
One of the fundamental issues in these studies is the observation of vortices with quantized circulation in response to rotation \cite{mcw00,hce01,hhh01,rav01,arv01,ech02,ech03,sce04}, which intrinsically related to the existence of superfluidity associated with BEC \cite{ps03}.
In experiments, vortices are produced using various techniques like stirring the condensate with a laser beam \cite{mah99,mcb01}, imprinting an appropriate phase pattern onto the condensate trapped in a Joffe-Pritchard magnetic trap \cite{lgc02} or rotating the condensate directly by mechanical means \cite{arv01}, etc.
The majority of experiments have focused on externally rotated condensates, where vortices of the same circulation sign arrange themselves into vortex lattices \cite{mcw00,rav01,arv01}.
Further developments have been focused on creating regular lattices with a number of singly quantized vortices \cite{rav01,arv01,ech02,ech03,sce04}, multiply quantized vortices \cite{lgc02,ssv04,nct13} and giant vortices \cite{ech03,cpr13}.
On the theoretical front, the rotational properties of BEC and creation of vortices in a harmonic trap have been analyzed mostly by the mean-field approach like Gross-Pitaevskii scheme as in Refs.~\cite{br99,kmp00,lf99,lnf01,gp01,vvh05,cpr13} or beyond the mean-field approximation \cite{kmp00,wg00,cwg01,pb01,ahs01,lhc01_pra,ryl04,blo06,rkym06,ryb06,un06,dbo07,dbld09,bg10,ckm13,ia17}.
To examine quantum mechanical phase coherence of the Bose-condensed gas, recent studies have been elucidated that the quantum entanglement in terms of von Neumann entropy, can be used as a probe to analyze the novel phases of many-body quantum states \cite{py01,lgc09,lf10,ecp10,ia16,ia20}.
A review of basic results on BEC vortices can be found in Ref.~\cite{bdz08,coo08,fet09,srh10,im17}.
\\
\indent
Our aim in the present work is to study the many-body quantum correlation in the harmonically trapped quasi-2D Bose-Einstein condensate, hosting upto seven vortices, with repulsive finite-range Gaussian interaction.
In order to analyze the ground state properties, exact diagonalization of the ${\bf n} \times {\bf n}$ Hamiltonian matrix has been performed in given subspaces of (quantized) total angular momentum with Hilbert dimensionality ${\bf n} \sim 10^{5}$ (e.g., for $L_{z}=N=16$, ${\bf n}=384559$).
In constructing the many-body basis states we go beyond the lowest Landau level approximation so as to include the single-particle basis states with radial quantum number $n_{r} \geq 0$ and angular momentum quantum number $m$ of either sign \cite{ahs01,lhc01_pra}.
To investigate the stability of the quantum states, the critical angular velocity of the rotating system is discussed.
The degree of condensation and von Neumann entanglement entropy of angular momentum states are calculated as a measure of many-body quantum correlation.
To analyze the internal structure of stable as well as unstable states, the conditional probability distribution is also obtained.
Beyond the mean-field approximation, we explore the novel phases of vortex states in subspaces of quantized total angular momentum and their transition to stable vortex patterns (configurations) is examined.
\\
\indent 
This paper is organized as follows.
In Sec.~\ref{model}, we present the theoretical model of a rotating Bose system that describe the many-body Hamiltonian with repulsive Gaussian interaction. 
In Sec.~\ref{results}, we present the numerical results for $N=16$ bosons and discuss the physical quantities of interest (which are experimentally accessible) to examine in detail the ground state properties of the system in quasi-2D harmonic trap. 
Finally in Sec.~\ref{conc}, we briefly summarize our results of the present study.
Few important supplemental derivations for exact diagonalization calculation are deferred to the Appendixes.
\section{MODEL}
\label{model}
We consider a bosonic system of $N$ identical atoms confined strongly in an external harmonic trap potential
$V({\bf r})={\frac{1}{2}}M\omega_{r}^{2}\left({r}^{2} + \lambda_{z}^{2} {z}^{2}\right)$ with anisotropic ratio $\lambda_{z}\equiv \omega_{z} / \omega_{r}\gg 1$, where $M$ is mass of atoms and $\omega_{r}~ (\omega_{z})$ is the radial (axial) frequency.
With strong axial confinement, our system is effectively quasi-two-dimension (quasi-2D) having $x$-$y$ symmetry, rotating about the $z$-axis with trap angular velocity $\Omega \equiv \widetilde{\Omega}/{\omega_{r}}$ $(\leq 1)$.
Introducing $\hbar \omega_{r}$ as the unit of energy and $a_{r} = \sqrt{\hbar/{M \omega_{r}}}$ as the corresponding unit length, the dimensionless Hamiltonian in the co-rotating frame is given as 
\begin{equation}
{H}^{rot}={H}^{lab}-\Omega L_{z}
\label{roth}
\end{equation}
where $L_{z}$ (scaled by $\hbar$) is the $z$ projection of the total angular momentum operator and
\begin{equation}
H^{lab} = \sum_{i=1}^{N} \left[-\frac{1}{2} \bm{\nabla}^{2}_{i} + \frac{1}{2} {\bf r}_{i}^{2} \right] + \frac{1}{2} \sum_{i\neq j}^{N} U \left({\bf r}_{i},{\bf r}_{j}\right)
\label{mbh}
\end{equation} 
The atom-atom interaction $U\left({\bf r}_{i},{\bf r}_{j}\right)$ is modelled by the Gaussian potential with parameter $\sigma$ (scaled by $a_{r}$) being the effective range of the potential \cite{cfa09,dka13,iasl15,ia15,im17,ia20}
\begin{equation}
U \left({\bf r}_{i},{\bf r}_{j}\right) = \frac{\mbox{g}_{2}} {2\pi{\sigma^{2}}}
\exp{\left[ -\frac{\left({\bf r}_{i}-{\bf r}_{j}\right)^{2}}{2\sigma^{2}} \right]}
\label{gip}
\end{equation}
where $\mbox{g}_{2}=4\pi {a_{s}}/{a_{r}}$ is a measure of strength of the two-body interaction with $a_{s}>0$ being the positive $s$-wave scattering length so that the effective interaction is repulsive \cite{ias98,cgj10}.
For a given value of $\sigma$, the $s$-wave scattering length in the interaction strength parameter $\mbox{g}_{2}$ is adjusted in such a way that the gas parameter $\left(Na_{s}/a_{r}\right)$ becomes relevant to the experimental value \cite{ahs01,dgp99}.
\section{NUMERICAL RESULTS AND DISCUSSION}\label{results}
The results are presented for an interacting system of $N=16$ Bose atoms of $^{87}$Rb confined in quasi-2D harmonic trap, with anisotropic ratio $\lambda_{z} \left(\equiv \omega_{z}/ {\omega_{r}}\right)=\sqrt{8}$ and radial frequency $\omega_{r}= 440\pi$ Hz corresponding to the trap length $a_{r}=\sqrt{{\hbar}/{M\omega_{r}}}=0.727 \mu$m \cite{bp96}.
Using Davidson iterative algorithm \cite{dav75}, exact diagonalization of the ${\bf n} \times {\bf n}$ Hamiltonian matrix is carried out beyond lowest Landau level approximation \cite{ahs01,lhc01_pra} separately for each of the subspaces of total angular momentum in the regime $0\leq {L}_{z} \leq 5N$.
For a many-body system (under consideration here), the characteristic energy scale for the interaction is determined by the dimensionless gas parameter $Na_{s}/a_{r}$, where $a_{s}$ is the tunable $s$-wave scattering length \cite{ias98,cgj10}. 
Owing to increasing dimensionality of the Hilbert space with increase in number of atoms $N$, the computation (exact diagonalization) becomes impractical beyond a few hundred atoms, for instance, ${\bf n}=384559$ for $L_{z}=N=16$.
We, therefore, vary $s$-wave scattering length $a_{s}$ in our calculations to achieve a suitable value of $\left(N a_{s}/a_{r}\right) \le 1$, relevant to experimental situation \cite{dgp99}.
Accordingly, the parameters of two-body Gaussian potential in Eq.~(\ref{gip}) have been chosen as: interaction range $\sigma = 0.1$ (in units of $a_{r}$) and interaction strength parameter $\mbox{g}_{2} \left(=4\pi {a_{s}}/{a_{r}}\right)= 0.9151$ corresponding to $a_{s}=1000a_{0}$, where $a_{0}=0.0529~nm$ is the Bohr radius.
\subsection{Stability of quantum states}\label{sec:cav}
We begin by examining how a quasi-2D Bose-condensed gas with repulsive finite-range Gaussian interaction responds to an externally impressed rotation in the angular momentum regime $0\le L_{z} \le 5N$, where $N$ is the number of bosons.
In the co-rotating frame, the simultaneous eigenstate of Hamiltonian (\ref{roth}) and total angular momentum minimizes the energy $\left\langle \Psi\left|\left({H^{lab}-\Omega L_{z}}\right)\right|\Psi\right\rangle $ at zero temperature, to become the ground state of the system.  
This is equivalent to minimizing $\left\langle H^{lab} \right\rangle$ with respect to $\Psi$, subject to the constraint that the system has angular momentum expectation value $L_{z}$ with angular velocity $\Omega$ identified as the corresponding Lagrange multiplier.
Here $\Psi$ is the variationally obtained $N$-body ground state associated with $L_{z}$.
Thereby for the stable states in the rotating frame, one may obtain a number of successive critical angular velocities $\Omega_{\bf c} \left(L_{z}^{i}\right);~i=1,2,3,\dots$ for the angular momentum states at which a particular excited state becomes the ground state of the system.
However for unstable angular momentum states such a critical angular velocity doesn't exist, which therefore remain as the excited states.
Thus, the critical angular velocity $\Omega_{\bf c}\left(L_{z}^{i}\right)$ is the one, beyond which the higher total angular momentum state $L_{z}^{i}$ becomes lower in energy in the rotating frame compared to the lower angular momentum state $L_{z}^{\left(i-1\right)} \left(<{L}_{z}^{i}\right)$ and is given by \cite{ahs01,im17}: 
\begin{equation}
\Omega_{\bf c}\left(L_{z}^{i}\right) = \frac{E^{lab} \left(L_{z}^{i}\right)-E^{lab}\left(L^{\left(i-1\right)}_{z}\right)}{L_{z}^{i}-L^{\left(i-1\right)}_{z}}
\label{cav}
\end{equation}
where $E^{lab}\left(L_{z}\right)$ is the variationally obtained $N$-body state energy for the total angular momentum state $L_{z}$ in the non-rotating frame. 
\begin{figure}[!htb]
\includegraphics[width=1.0\linewidth]{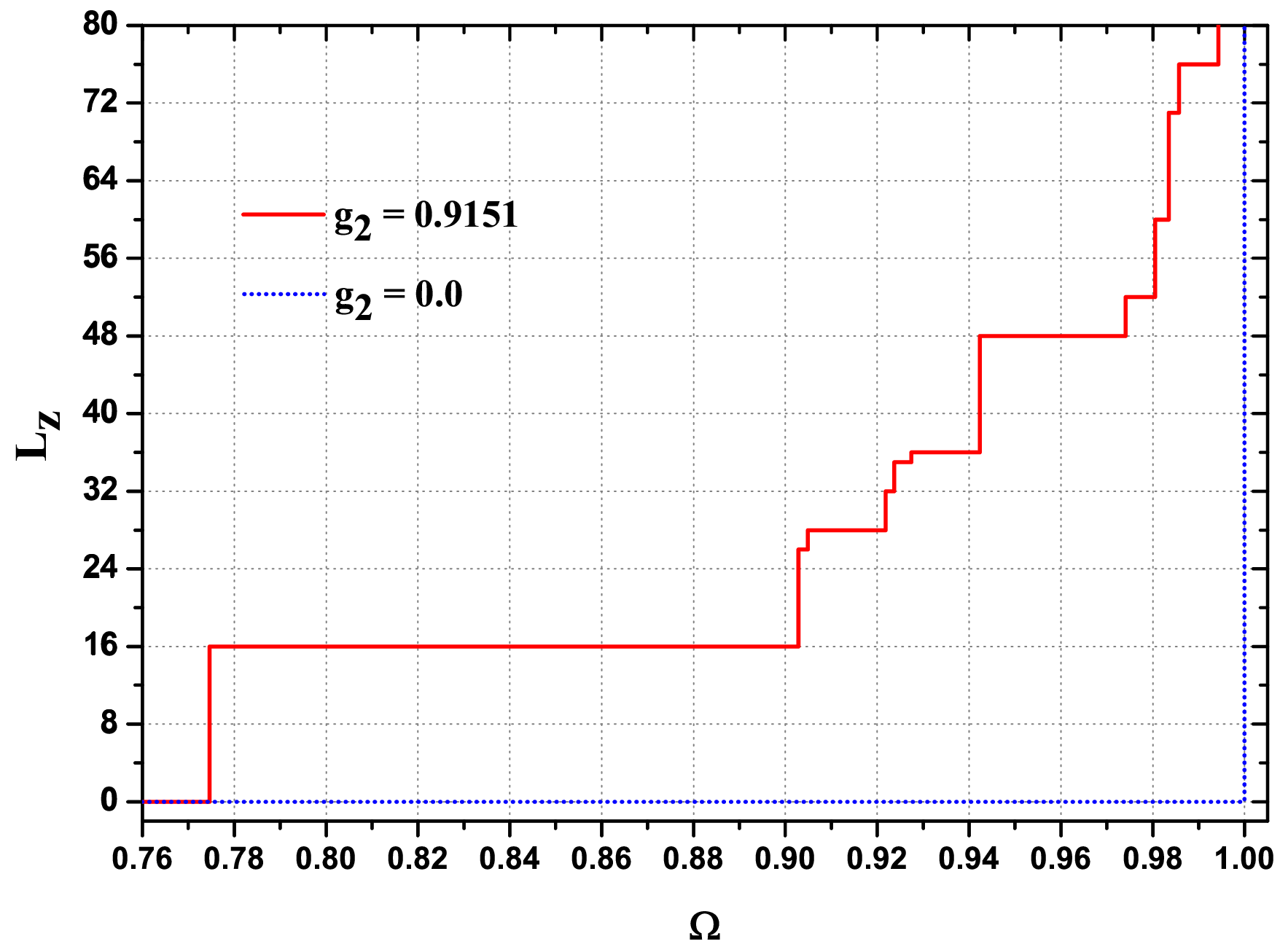}
\caption{\label{fig:stbl} Variation of the total angular momentum $L_{z}$ with the scaled angular velocity $\Omega$ for a system of rotating Bose-condensed gas of $N=16$ particles interacting via finite-range Gaussian potential (solid lines). The interaction parameters have been taken as $\mbox{g}_{2}=0.9151$ and $\sigma=0.1$ in Eq.~(\ref{gip}). The vertical lines represent quantum phase transition between stable vortex states of different symmetries. The horizontal plateaus represent the stability of the corresponding angular momentum state with respect to rotation $\Omega$. The Dashed line represent the non-interacting case, included here for reference.}
\end{figure}
\\
\indent
Fig.~\ref{fig:stbl} shows the total angular momentum $ L_{z}$  {\em versus} the angular velocity $\Omega $, for the many-body ground state, and is called the $L_{z}-\Omega$ stability curve.
There exist a series of critical angular velocities $\Omega_{\bf c} \left(L_{z}^{i}\right) <1;~~i=1,2,3,\dots$, at which the ground state of the rotating Bose-condensed system changes its quantized total angular momentum abruptly.
For $N=16$ bosons with Gaussian repulsion~(\ref{gip}), we observe a step-like structure in the $L_{z}-\Omega$ stability curve (Fig.~\ref{fig:stbl}) with plateaus at $L_{z}^{i}=$ 0, 16, 26, 28, 32, 35, 36, 48, 52, 60, 71, 76.
These $L_{z}^{i}$ ground states corresponding to critical angular velocities $\Omega_{\bf c}\left(L_{z}^{i}\right)$, are referred to as quantum mechanically stable states \cite{lf99,lnf01,gp01}.
Thus, for rotating bosons interacting via finite-range Gaussian potential, one may recovers the step-like structure in the stability curve, well known for the Bose-condensed gas interacting via zero-range ($\delta$-function) potential \cite{coo08}.
\\
\indent
Initially, when the Bose-condensed system is non-rotating, the angular momentum state $L_{z}=0$ corresponds to the ground state of the system. 
As the system is subjected to an external rotation, higher angular momentum states $\left(L_{z} > 0\right)$, which minimize the free energy, become the ground state of the system.
As observed in Fig.\ref{fig:stbl}, the state with total angular momentum $L_{z}=0$ continues to be the ground state of the system till angular velocity  $\Omega_{{\bf c}}\left(L_{z}^{1}=N\right)$, where the state with total angular momentum $L_{z}=N$, becomes the new ground state in the co-rotating frame. 
The system thereby undergoes a quantum phase transition from a non-rotating state to the single vortex state, seen as vertical line (a ``jump" in ground state total angular momentum) in the stability curve.
At angular velocities beyond $\Omega_{{\bf c}}\left(L_{z}^{1}=N\right)$, the ground state angular momentum has a plateau $L_{z}=N$ till angular velocity reaches $\Omega_{{\bf c}}\left(L_{z}^{2}\right)$, where the second jump in ground state angular momentum takes place. 
Thereafter, one observes a sequence of plateaus and jumps.
The rotational angular velocity $\Omega~(\equiv\widetilde{\Omega}/{\omega_{r}})$ grows from lowest limit zero to the upper limit $\Omega=1$ beyond which the condensate will become centrifugally unstable. 
We notice that with increasing $\Omega<1$, the width of the plateaus and the hight of the jumps in the stability curve, in general, decreases which may plausibly be attributed to condensate depletion, as discussed later in the section.
\\
\indent
The critical angular velocities $\left\{ \Omega_{{\bf c}}\left(L_{z}^{i}\right)\right\}$ are calculated for the onset of different vortex states and undergoes quantum phase transition.
At an angular velocity $\Omega < \Omega_{\bf c}\left(L_{z}^{i}\right)$, the nucleation of the first  vortex state with angular momentum $L_{z}^{i}=16$ begins which stabilizes at $\Omega =\Omega_{\bf c}\left(L_{z}^{i}\right)$ and becomes the new ground state of the system. 
For  angular velocities beyond $\Omega_{\bf c}\left(L_{z}^{i}\right)$, the vortex state $L_{z}^{i}$ continues to be the ground state of the system till the nucleation of the next vortex state begins. 
The angular momentum states $L_{z}^{i} = 26,28,32,36,48,52$, correspond to off-centred vortex states (without a central vortex), as will be seen when we examine their internal structure in later sections.
The single-particle angular momentum $\ell_{z}=L_{z}/N$ is not an integer multiple of $\hbar$.
In contrast, for axisymmetric multi-vortex states, the total angular momentum $L_{z}$ is quantized in integer units and hence the topological discreteness of the vortices.
We next correlate these observations with our results on the many-body ground states for their quantum correlation in terms of von Nuemann entropy as well as degree of condensation.
\subsection{Many-body quantum correlation}
\begin{figure}[!htb]
\includegraphics[width=1.0\linewidth]{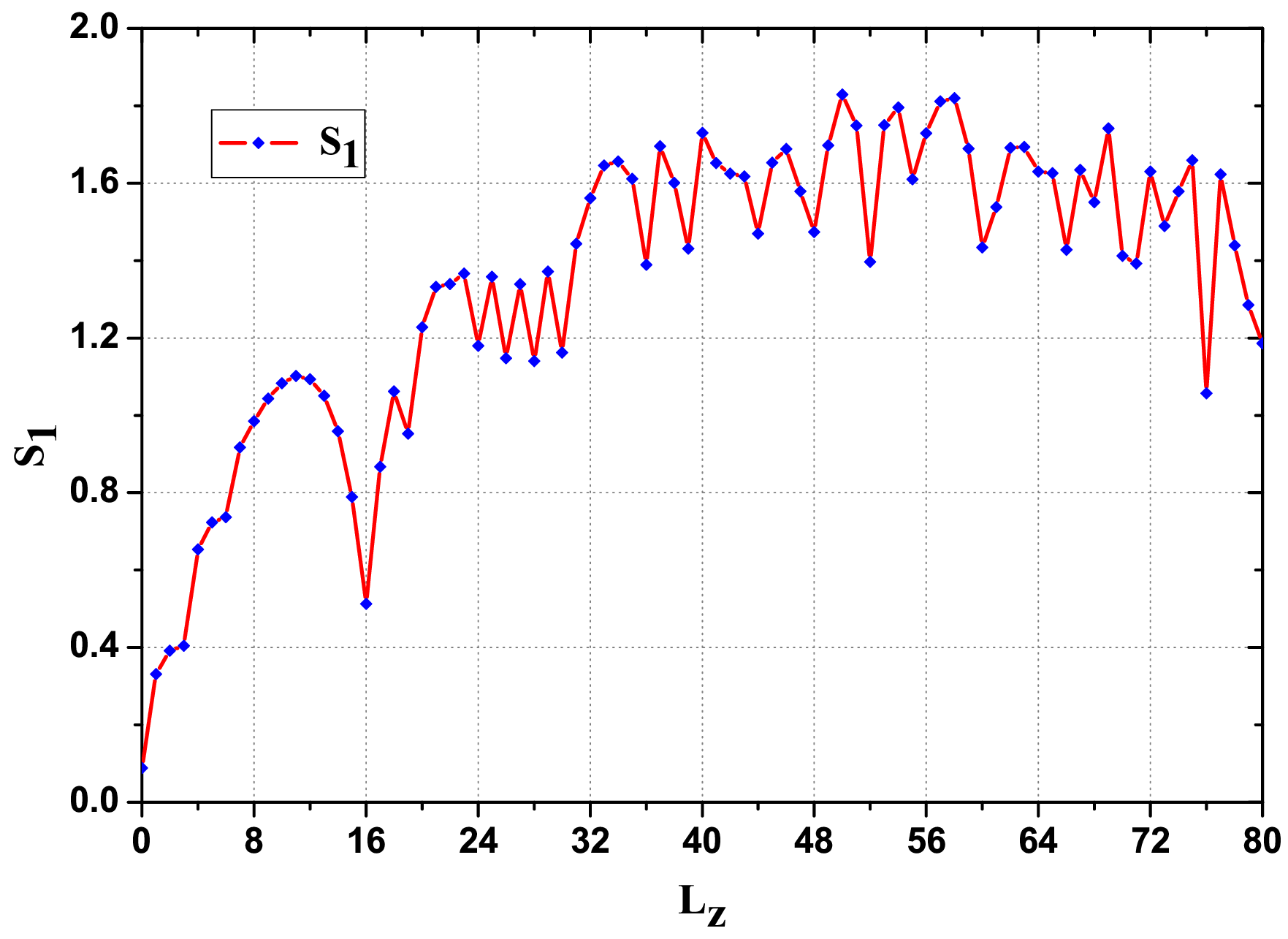} 
\caption{\label{fig:n16qent1} For a rotating system of $N=16$ bosons, the von Neumann entropy $S_{1}$ of the ground states in the subspaces of total angular momentum $L_{z}$ with interaction strength $\mbox{g}_{2}=0.9151$ and range $\sigma=0.1$ of the Gaussian potential~(\ref{gip}).}
\end{figure}

An interesting and useful measure of the quantum phase correlation in the many-body ground state of a confined system is provided by the von Neumann entanglement entropy \cite{py01,lgc09,lf10,ecp10,ia16,ia20}, an extension to the classical Gibbs entropy, defined as
\begin{equation}
S_{1} = -\mbox{Tr}\left(\hat{\rho} \ln \hat{\rho}\right)
\end{equation}
where $\hat{\rho}$ is the reduced single-particle density operator obtained from the many-body ground state wavefunction. 
In terms of eigenvalues $\left\{\lambda_{\mu }\right\}$ and eigenfunctions $\left\{ \chi_{\mu }\left({\bf r}\right)\right\}$, the reduced single-particle density operator $\hat{\rho}$ takes the form as in Eq.~(\ref{spd}), for details please see Appendix \ref{sec:rspdm}.
The von Neumann entropy is, then, evaluated explicitly as
\begin{equation}
S_{1} = -\sum_{\mu}\lambda_{\mu}\ln{\lambda_{\mu}}
\label{vent}
\end{equation}
in different subspaces of quantized total angular momentum $L_{z}$, as shown in Fig.~\ref{fig:n16qent1} for $N=16$ bosons.
\begin{figure}[!htb]
\includegraphics[width=1.0\linewidth]{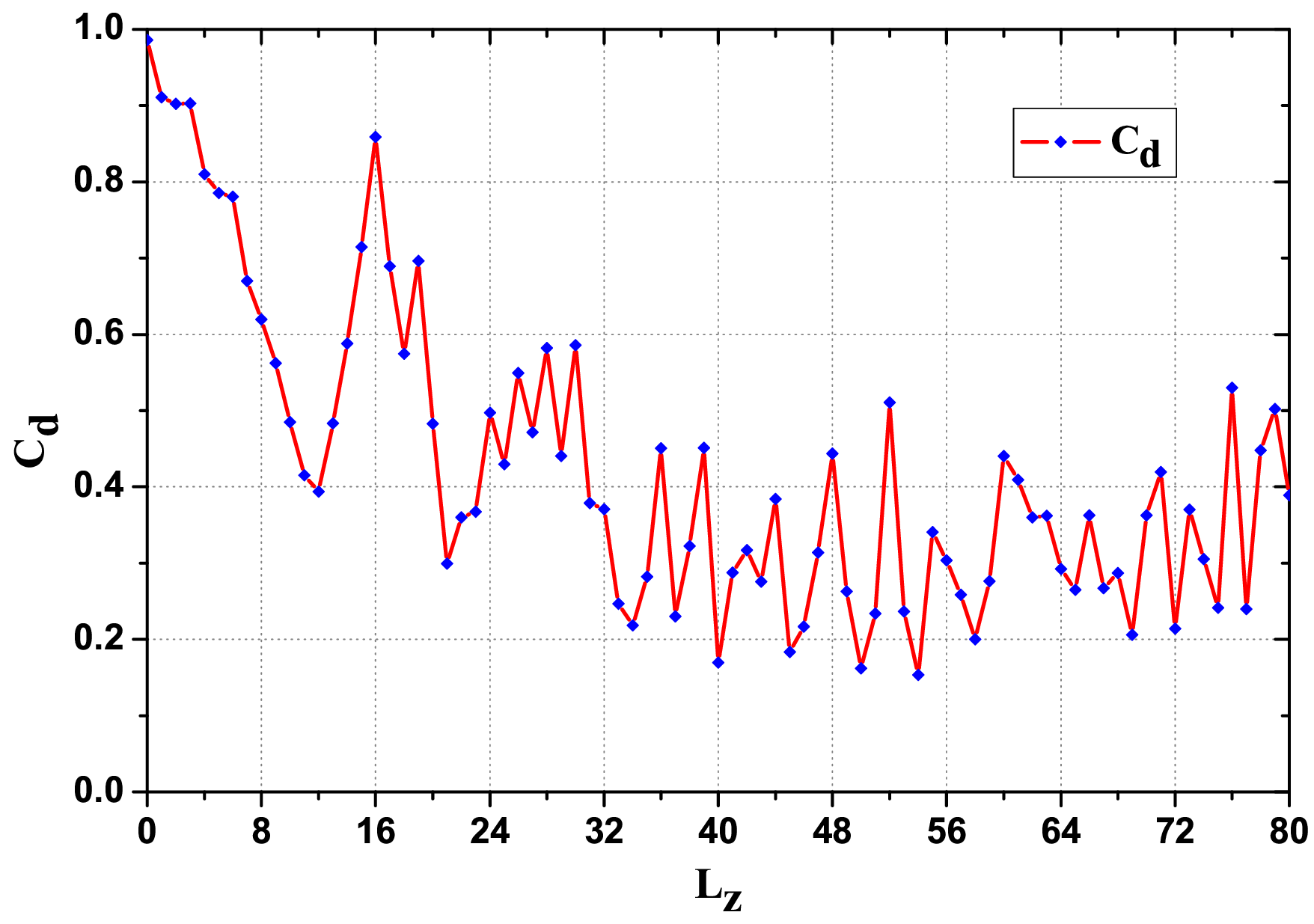} 
\caption{\label{fig:n16doc} The degree of condensation ($C_{d}$) as a function of the total angular momentum $L_{z}$ for a rotating system of $N=16$ bosons with interaction strength $\mbox{g}_{2}=0.9151$ and range $\sigma=0.1$ of the Gaussian potential~(\ref{gip}).}
\end{figure}
\\
\indent
It is clear from Fig.~\ref{fig:n16qent1} that for a (harmonically confined) rotating Bose gas, the von Neumann entropy $S_{1}$ has a tendency to grow with angular momentum $L_{z}$, but the whole curve shows a slightly oscillatory behavior.
Further, the curve of entropy $S_{1}$ {\it versus} $L_{z}$ exhibits a series of local minima $L_{z}^{S_{1}}$.
For instance, the local minima appear at angular momenta $L_{z}^{S_{1}}=$ 0, 16, {24}, 26, 28, {30}, 36, {39}, {44}, 48, 52, {55}, 60, {66}, 71, 76 as shown in Fig.~\ref{fig:n16qent1}. 
It is interesting to note that the entropy $S_{1}$ plotted in Fig.~\ref{fig:n16qent1}, present features similar to that of stability curve in Fig.~\ref{fig:stbl}. 
In fact, most of these local minima (obtained from $S_{1}$) correspond to the stable angular momentum $L_{z}^{i}$-states of ground state $\Psi$, characterized by critical angular velocities $\Omega_{{\bf c}}\left(L_{z}^{i}\right);\ i=1,2,3\dots$ (see Fig.~\ref{fig:stbl}).
Therefore, the von Neumann entropy $S_{1}$ plotted in Fig.~\ref{fig:n16qent1}, can also be used to study the $L_{z}$-$\Omega$ stability curve Fig.~\ref{fig:stbl}, 
where the plateaux are found at the angular momentum values $L_{z}^{i}=$ 0, 16, 26, 28, {32}, {35},  36, 48, 52, 60, 71, 76.
Similar results are shown in \cite{lgc09}, for zero-range ($\delta$-function) interaction potential in LLL approximation.
\\
\indent
The von Neumann entanglement entropy $S_{1}$ defined in Eq.~(\ref{vent}) also provides information about the degree of condensation ${C}_{d}$ defined in Eq.~(\ref{doc}), which is sensitive to the loss of macro-occupation, see Appendix \ref{sec:doc}.
To ascertain the relation between ${C}_{d}$ and $S_{1}$, we compare corresponding plots of both these quantities with respect to angular momentum $L_{z}$, as shown in Fig.~\ref{fig:n16doc} and \ref{fig:n16qent1}, respectively.
It is observed from Figures that the value of $C_{d}$ approaches to unity and the value of entropy $S_{1}$ reduces to zero, for a perfect Bose-Einstein condensate \cite{zhs08}, since all bosons occupy one and the same mode {\em i.e.} $\lambda_{1}=1$ and $\lambda_{\mu}=0$ for $\mu=2,3,\dots$.
The angular momentum state $L_{z}=0$ corresponds to a fairly condensed state with $S_{1}\sim 0$.
As the condensate depletes, with more than one eigenvalue $\left\{ \lambda_{\mu} \right\}$ becoming non-zero, the entropy $S_{1}$ increases.
We observe that lesser the value of $S_{1}$, greater will be the quantum many-body correlation between the particles thus enhancing the condensation.
In the single vortex state corresponding to $L_{z}=N=16$, the degree of condensation $C_{d}$ has a larger value and there is a dip in the entropy $S_{1}$. 
For other vortex states too, we have found the dip in $S_{1}$ and corresponding peak in $C_{d}$ at the stable angular momentum states associated with the critical angular velocities of the system. 
\begin{table}[!htb]
\caption{\label{tab:n16_table1} The many-body ground state energies $E^{lab}(L_{z}^{i})$ in the laboratory frame, the values of critical angular velocity $\Omega_{{\bf c}}\left(L_{z}^{i}\right);i=1,2,3,\dots$ and the $p$-fold rotational symmetry of each stable vortex state for $N=16$ bosons in the given subspaces of $0 \le L_{z} \le 5N$. The single-particle quantum number $m_{1}$ corresponding to largest eigenvalue $\lambda_{1}$ of the reduced single-particle density matrix in Eq.~(\ref{spd}), the degree of condensation (${C}_{d}$) and von Neumann entropy ($S_{1}$) is also reported. Here, the interaction strength $\mbox{g}_{2}=0.9151$ and range $\sigma=0.1$ of the Gaussian potential~(\ref{gip}).}
\begin{ruledtabular}
\begin{tabular}{ccccccc}
$L_{z}^{i}$ & $E^{lab}\left(L_{z}^{i}\right)$ & $\Omega_{{\bf c}}\left(L_{z}^{i}\right)$ & $p$ & $m_{1}$ & ${C}_{d}$ & $S_{1}$ \\ 
\hline 
 0 & 47.09788 & 0.0 & - & 0 & 0.9857 & 0.0890 \\ 
16 & 59.49139 & 0.7745 & - & 1 & 0.8589 & 0.5128 \\
26 & 68.52023 & 0.9028 & 2 & 2 & 0.5493 & 1.1478 \\
28 & 70.32997 & 0.9048 & 2 & 2 & 0.5823 & 1.1402 \\ 
32 & 74.01745 & 0.9218 & 2 & 2 & 0.3707 & 1.5611 \\
35 & 76.78844 & 0.9236 & 3 & 3 & 0.2821 & 1.6116 \\
36 & 77.71583 & 0.9273 & 3 & 3 & 0.4506 & 1.3889 \\
48 & 89.02375 & 0.9423 & 4 & 4 & 0.4437 & 1.4733 \\
52 & 92.92004 & 0.9740 & 4 & 4 & 0.5105 & 1.3965 \\
60 & 100.76423& 0.9805 & 5 & 5 & 0.4407 & 1.4345 \\ 
71 & 111.58246& 0.9835 & 5 & 6 & 0.4197 & 1.3919 \\
76 & 116.51123& 0.9857 & 6 & 7 & 0.5299 & 1.0564 \\
\end{tabular}
\end{ruledtabular}
\end{table}
\\
\indent
For repulsive Bose-condensed gas rotating with angular velocity $\Omega_{{\bf c}}\left(L_{z}^{i}\right)$ smaller than the trapping frequency $\omega_{r}$, there exists stable states containing the various vortex configurations.
The system corresponding to these stable vortex states is well described by a stationary state with some finite non-zero vorticity $m_{1}$, identified as the angular momentum quantum number of the most dominant single-particle state $\chi_{1}(r)$ corresponding to the largest condensate fraction $\lambda_{1}$ (macroscopic eigenvalue) of the reduced single-particle density matrix (RSPDM) defined in Eq.~(\ref{spd2}).
For each stable vortex state in the regime $0\le L_{z}\le 5N$ with $N=16$ bosons, Table~\ref{tab:n16_table1} list values of the following: (i) the lowest energy $E^{lab}(L_{z}^{i})$ of the $N$-body states in the laboratory frame (ii) critical angular velocity $\Omega_{\bf c}\left(L_{z}^{i}\right);i=1,2,3,\dots$ (iii) respective discrete rotational symmetry of the vortex states denoted by integer $p$, (iv) along with the vorticity defined by single-particle angular momentum quantum number $m_{1}$ of the RSPDM, (v) the degree of condensation ${C}_{d}$ and (vi) the von Neumann entanglement entropy $S_{1}$ of harmonically confined rotating Bose-condensate, corresponding to the stable states as observed in $L_{z}$-$\Omega$ stability curve (see Fig.~\ref{fig:stbl}).
In order to gain an insight we examine the spatial correlation by looking into the internal structure of the condensate for various stable as well as unstable vortex states.
\subsection{Internal structure of vortex states} \label{sec:cpd}
Circular symmetries of the rotating Hamiltonian often hide the internal structure in the exact many-body state. 
The pair correlation function, a standard tool in many-body physics, is often used to analyze the internal structure of the many-body state.
Here, the pair correlation function is conditional probability distribution (CPD) which provides a test for the presence of spatial correlation among bosons.
The CPD is defined as $\mathcal{P}\left({\bf r},{\bf r}_{0}\right)$ the probability of finding a particle at ${\bf r}$, given the presence of another at ${\bf r}_{0}$ and is computed by the expression \cite{yl00,lhc01_pra,blo06,ryl08,ckm13,ia17}: 
\begin{equation}
\mathcal{P} \left({\bf r},{\bf r}_{0}\right) = \frac{\langle \Psi\vert \sum_{i \neq j} \delta \left({\bf r}- {\bf r}_{i} \right)\delta \left({\bf r}_{0}-{\bf r}_{j}\right) \vert \Psi \rangle}{\left(N-1\right)\langle \Psi\vert \sum_{j} \delta \left({\bf r}_{0}- {\bf r}_{j} \right)\vert \Psi \rangle}
\label{cpd}
\end{equation}
Here $\vert \Psi_{}\rangle$ is the many-body ground state obtained through exact diagonalization.
The details of CPD in finite system are sensitive to the selection of reference point ${\bf r}_{0}=(x_{0},y_{0})$ \cite{krm02}. 
Usually, the ${\bf r}_{0}$ is given a value of the order of spatial extent of the confined system, so that the effects of pair repulsion between atoms do not affect the vortex structure significantly.
For large bosonic systems, CPD show the vortices clearly by giving ${\bf r}_{0}$ a significantly large value.
However in case of systems with small number of particles $N$ (or depleted condensate); in order to obtain optimum information, the reference point ${\bf r}_{0}$ is taken as a position where the density of the system is maximum. 
The reference point ${\bf r}_{0}$ in our calculation is taken to be located on the $x$-axis {\it i.e.} ${\bf r}_{0}=\left(r_{0},0\right)$ for a system of rotating Bose-condensed gas of $N=16$ particles.
For slow rotating regime ($0 \leq L_{z}\leq 2N$) where the condensate is large, we choose ${\bf r}_{0}=(3,0)$. However for moderately to rapidly rotating regime ($2N < L_{z} \leq 5N$) where the condensate depletes due to centrifugal effect, we choose ${\bf r}_{0}=\left(1.5,0\right)$.
We have carefully selected the position of reference point ${\bf r}_{0}$ to ensure that it never coincides with the vortex position (low density region).
It is observed that a small displacement of the reference point ${\bf r}_{0}$ alters the pair correlation considerably for unstable states.
The reference point thus acts as an infinitesimal (spontaneous) symmetry breaking perturbation on the two-body correlation function.
\begin{figure}[!htb]
{\includegraphics[width=0.325\linewidth]{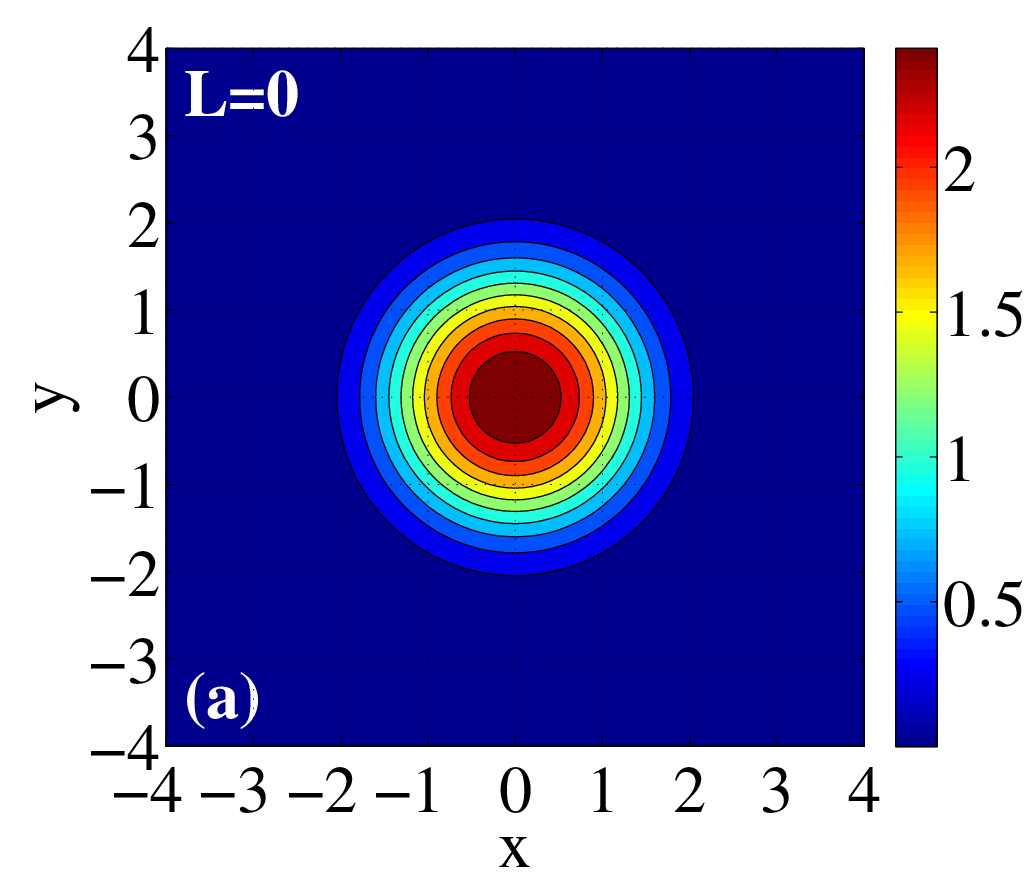}}
{\includegraphics[width=0.325\linewidth]{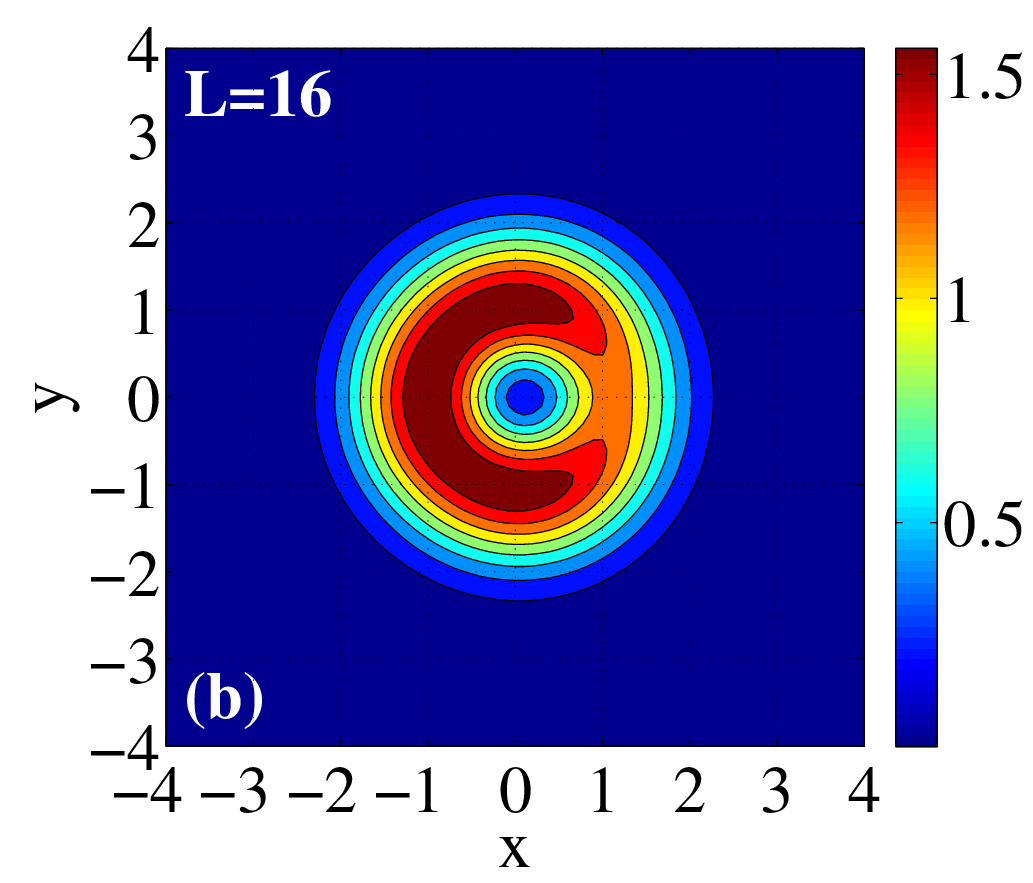}}
{\includegraphics[width=0.325\linewidth]{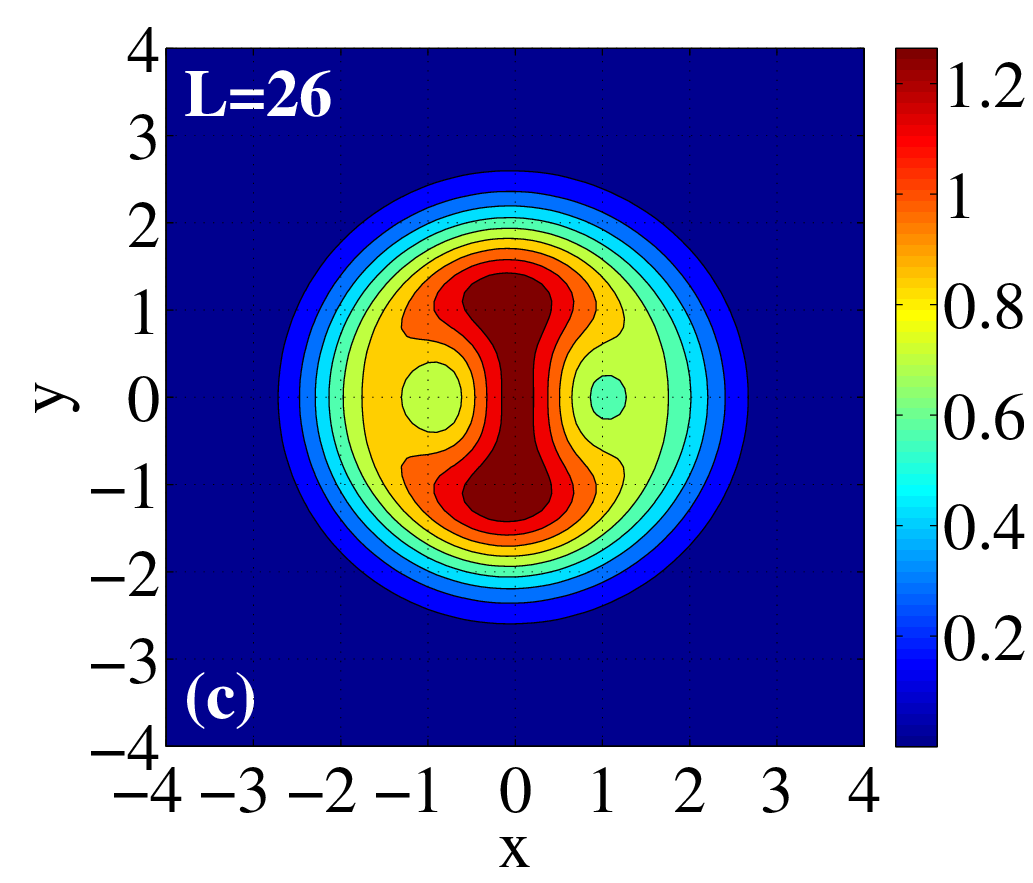}}
{\includegraphics[width=0.325\linewidth]{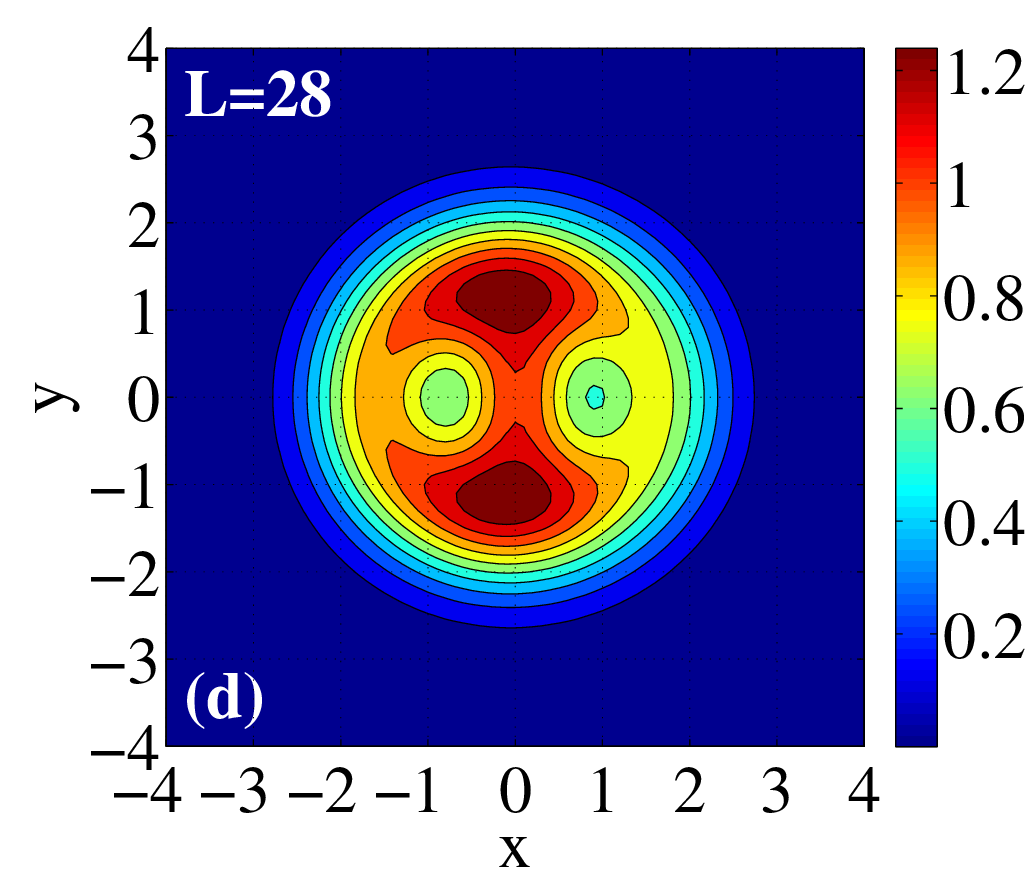}}
{\includegraphics[width=0.325\linewidth]{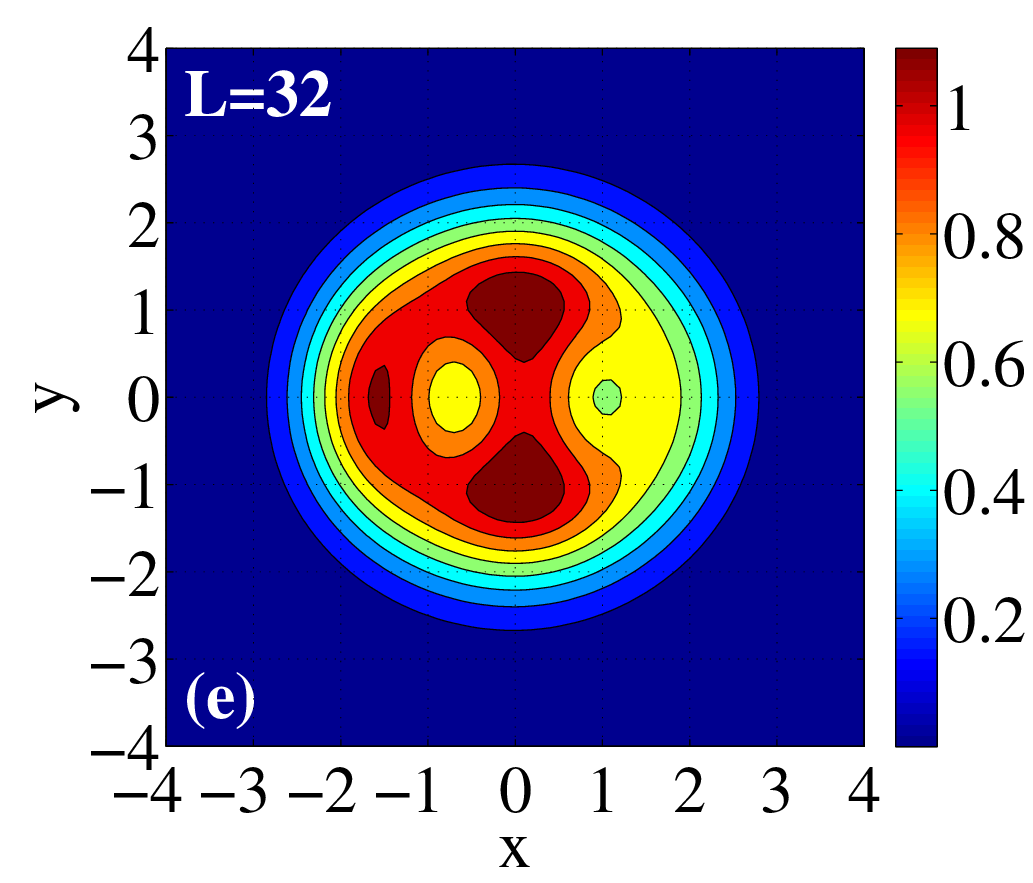}}
{\includegraphics[width=0.325\linewidth]{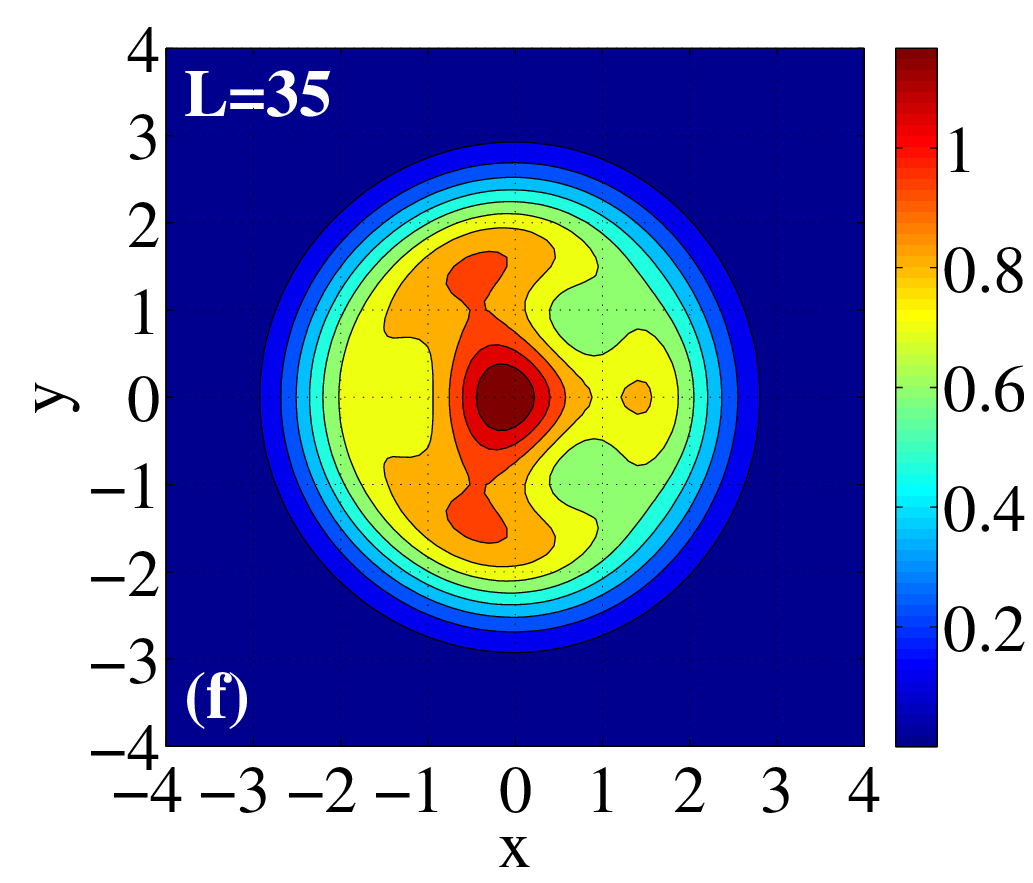}}
{\includegraphics[width=0.325\linewidth]{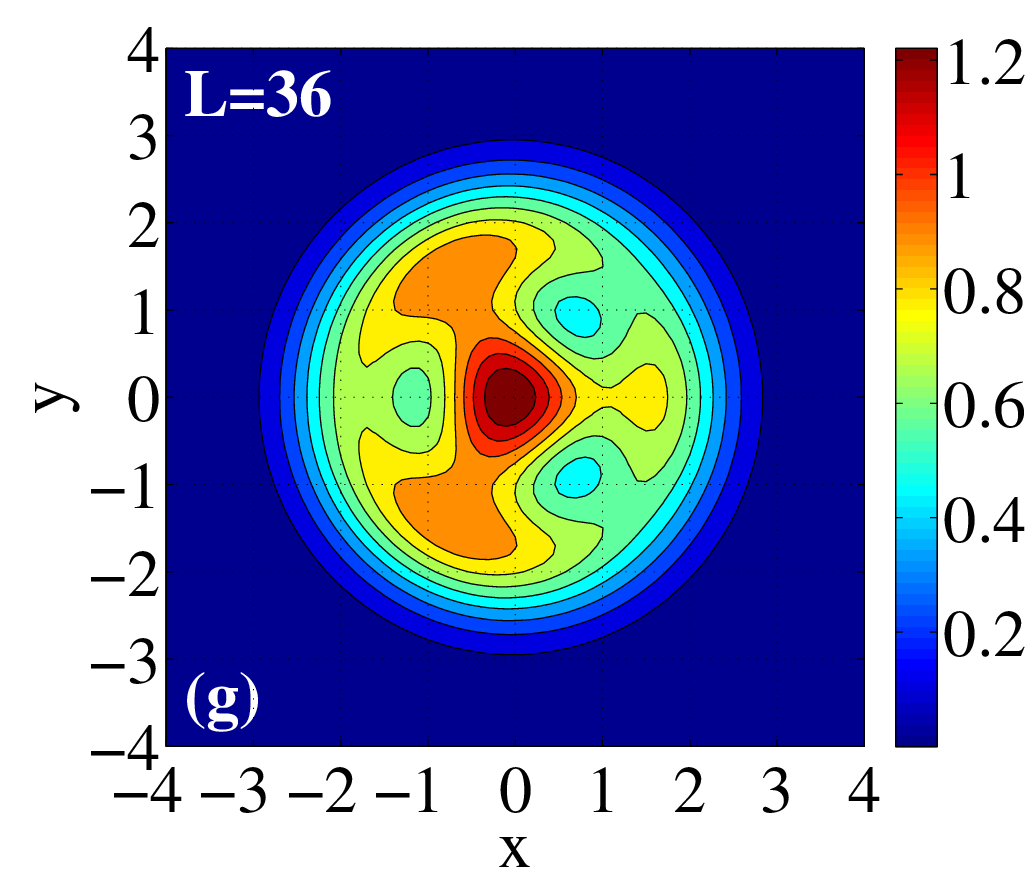}}
{\includegraphics[width=0.325\linewidth]{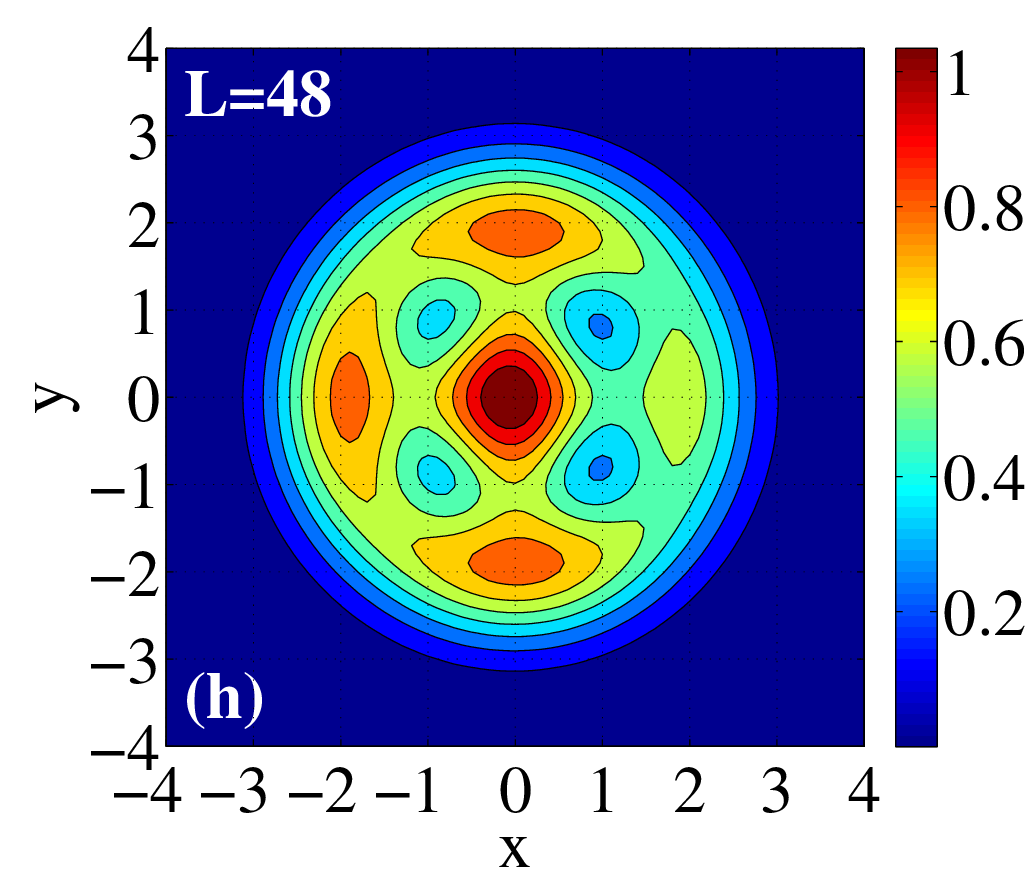}}
{\includegraphics[width=0.325\linewidth]{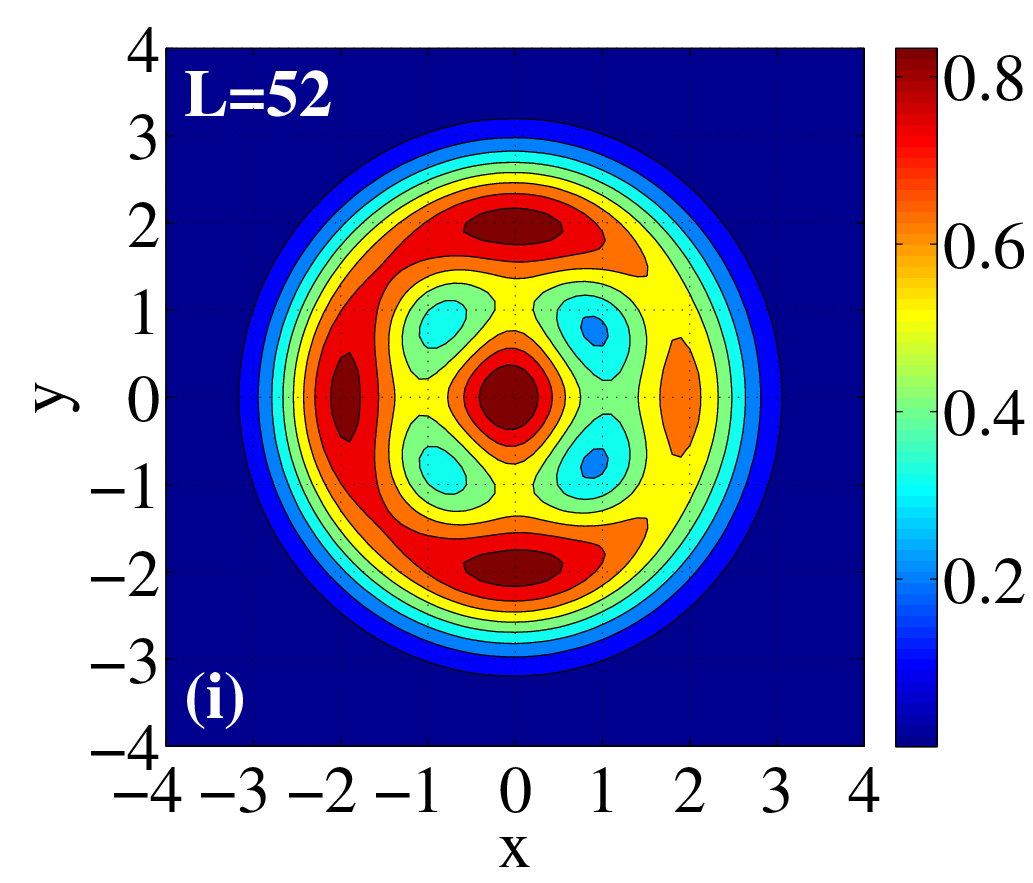}}
{\includegraphics[width=0.325\linewidth]{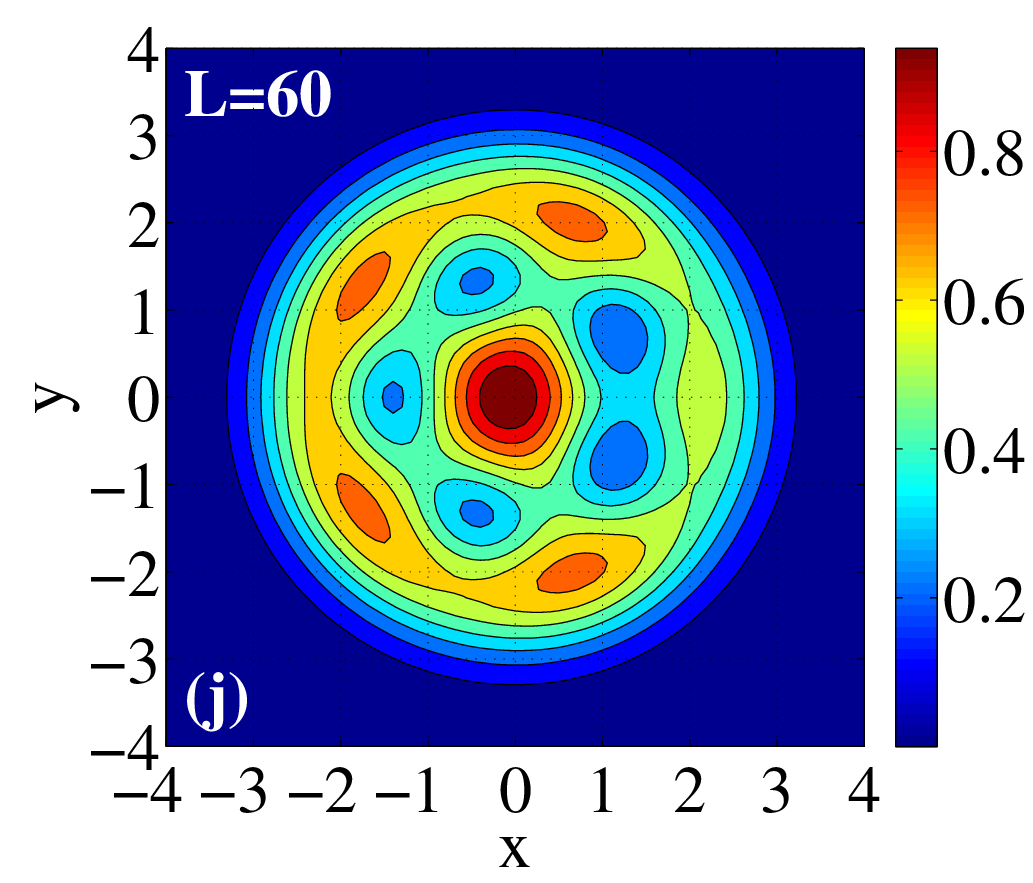}}
{\includegraphics[width=0.325\linewidth]{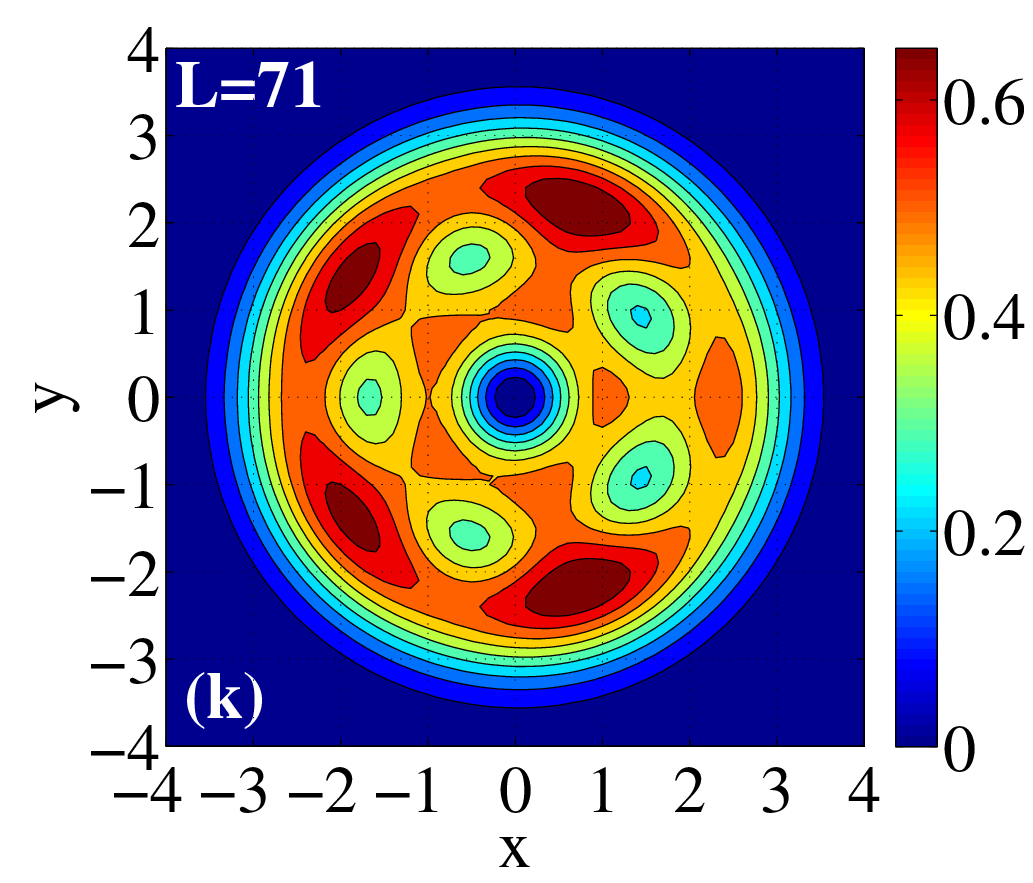}}
{\includegraphics[width=0.325\linewidth]{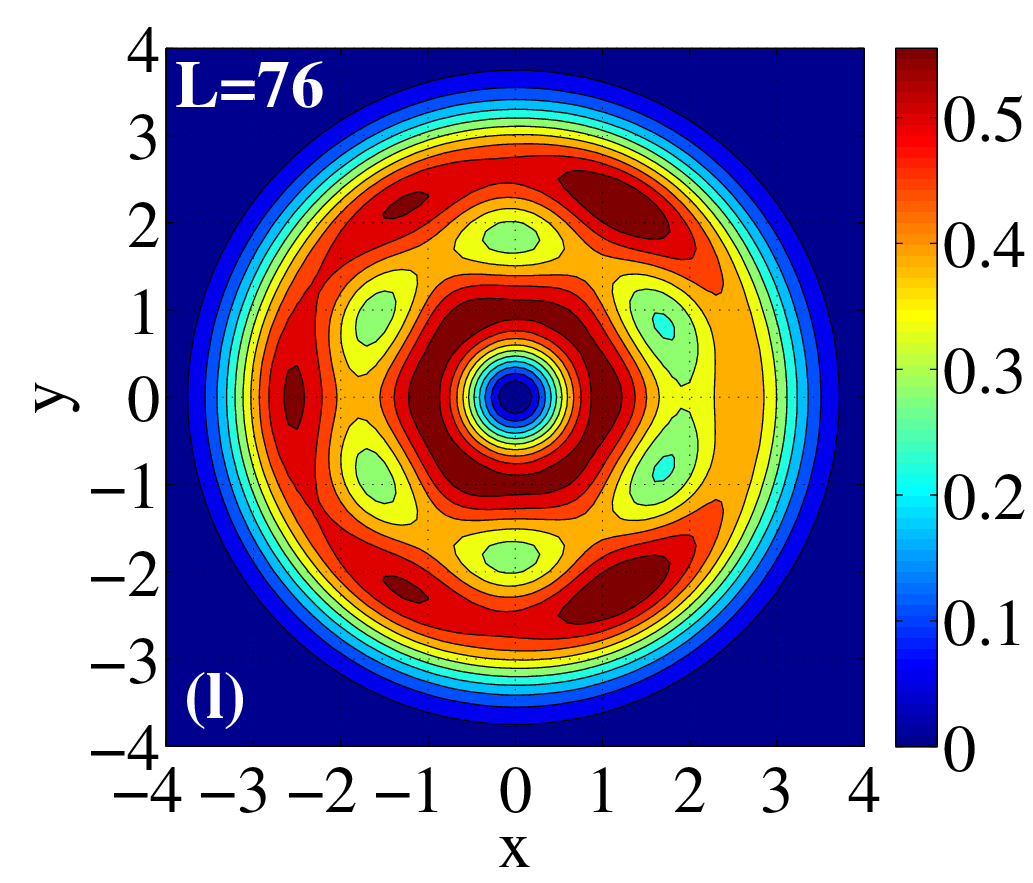}}
\caption{\label{fig:n16cpd}(Color online) CPD contour plots of stable vortex ground state for $N=16$ bosons in the rotating harmonic trap, with interaction parameter $\mbox{g}_{2}=0.9151$ and range $\sigma=0.1$ of the repulsive Gaussian potential~(\ref{gip}). Viewed along the axis of rotation ($z$-axis), each plot depicts an isosurface density profile which reflects the intrinsic distribution of bosons in the condensate for every successive critical angular velocity $\Omega_{\bf c}$ corresponding to stable vortex state, see Table~\ref{tab:n16_table1}. On the color bar, the dark-red region represents highest probability density falling off to blue region of lowest probability density. The reference point is located at ${\bf r}_{0}=(3,0)$ for vortex states $(a)$-$(e)$ and at $(1.5,0)$ for vortex states $(f)$-$(l)$.}
\end{figure}
\\
\indent
We have reported here the calculation for quantum mechanically stable ground states in the rotating frame (corresponding to critical angular velocities $\Omega_{{\bf c}}\left(L_{z}^{i}\right),~i=1,2,3,\dots$) for harmonically confined rotating Bose gas in the regime of $0 \leq L_{z} \leq 5N$.
It appears that $N = 16$, is sufficiently large system to give insight into vortex nucleation.
Calculating the CPD using Eq.(\ref{cpd}) for quantized angular momentum $L_{z}$-values, one may study the internal structure (spatial correlation) of $N$-body quantum states.
CPD is indeed an observable quantity in experiments which indicates the tendency of a system's evolution. 
Unlike the usual density distribution, which is circularly/cylindrically symmetric under rotational invariant confinement, the CPD plots exhibit asymmetric behavior and  reflect the intrinsic density distribution of particles for each successive critical angular velocity $\Omega_{{\bf c}}\left(L_{z}^{i}\right)$.
The respective CPD plots for $N = 16$ are presented in Fig.~\ref{fig:n16cpd} which depicts the condensate density isosurface, viewed along the axis of rotation ($z$-axis). 
The value of quantized total angular momentum $L_{z}$ is marked in the upper left corner of each CPD plot.
Knowing the stability of vortex states from $L_{z}$-$\Omega$ stability curve (see Fig.~\ref{fig:stbl}), we can extract valuable information about the vortex patterns (configurations) and their formation.
The angular momentum values associated with the stable ground states, accommodating 0, 1, 2, 2, 2, 3, 3, 4, 4, 5, 6, 7 vortices, are $L_{z}^{i}=$ 0, 16, 26, 28, 32, 35, 36, 48, 52, 60, 71, 76, respectively, as shown in Figs.~\ref{fig:n16cpd}(a)-\ref{fig:n16cpd}(l).
For an interacting Bose-condensed gas of $N=16$ particles, the different quantized angular momentum $L_{z}$ regimes observed may be listed as under:
(i) Fig.~\ref{fig:n16cpd}(a) for vortex-less non-rotating ground state with $L_{z}=0$,
(ii) Fig. \ref{fig:n16cpd}(b) for single central vortex state with $L_{z}=N=16$,
(iii) Figs.~\ref{fig:n16cpd}(c)-\ref{fig:n16cpd}(j) for multi-vortex state without a central vortex with $N < L_{z} < 4N$ and,
(iv) Figs.~\ref{fig:n16cpd}(k)-\ref{fig:n16cpd}(l) for polygonal vortex patterns having a central vortex with $4N \leq L_{z} \leq 5N$.
Thereby CPD plots show the transitions between stable states with different vortex patterns.
\\
\indent
A striking feature of the rotating Bose-condensate is their lack of full rotational symmetry. 
Instead they have only $p$-fold symmetry with $p=$ 2, 3, 4, 5 and 6 as shown in CPD plots of Fig.~\ref{fig:n16cpd}. 
Thus CPD can also be regarded as a measure of possible spontaneous symmetry breaking. 
A succession of discontinuous, symmetry-changing transitions are followed at higher values of angular momenta $L_{z}$, with each new stable state corresponding to a different configuration of vortices.
To elucidate, we have also produced a movie featuring the CPD contours of many-body states which provide information about evolution of the rotating finite system with different values of $L_{z}$ \emph{viz.} the formation of various pattern of vortices along with its discrete $p$-fold rotational symmetry (see the Supplemental Material \cite{video}).
The gradual entry of vortex in the rotating system can clearly be seen, first changing to a central single-vortex state at $L_{z}=N=16$, and then to higher vortex states.
When the angular momentum is increased beyond $L_{z}>N$, additional vortices appear in the form of regular structures as shown in Fig.~\ref{fig:n16cpd}.
As their number grows they may organise themselves in a triangular Abrikosov lattice \cite{arv01,ech02,tap13}.
\\
\indent
At $L_{z}$= 26, 28 and 32, the second vortex is appeared in the confined system as single-particle quantum number $m_{1}=1$ (vorticity associated with first centred vortex state) changes to $m_{1}=2$ with the formation of 2-fold symmetry vortex pattern. 
Out of these 2-fold vortex states, observed in $L_{z}-\Omega$ stability curve Fig. (\ref{fig:stbl}), the $L_{z}=28$ state with larger plateau size found to be more stable phase coherent state corroborated by the lower value of $S_{1}$ and higher value of $C_{d}$.
We further observe that in the angular momentum regime $22 \leq L_{z} \leq 63$ for $N=16$, no central vortex is formed along the axis of the trap as seen in Fig.~\ref{fig:n16cpd} (also refer to the \cite{video}).
It may be noted that the stable angular momentum state $L_{z}=60$ corresponds to 5-fold rotational symmetry of pentagonal vortex pattern without a central vortex at the trap center.
On the other hand the $L_{z}= 71$ and $76$ represent the 5-fold and 6-fold rotational symmetry respectively, of pentagonal and hexagonal vortex patterns with a central vortex.
Unlike the 6-fold rotational symmetry of hexagonal vortex pattern identified as a precursor to thermodynamically stable triangular vortex lattice, the vortex pattern with 5-fold rotational symmetry found to exist in confined system will not survive in the thermodynamic limit.  
Looking beyond the mean-field approximation it is important to mention here that, though we do not observe the linear configuration of three vortices as suggested in \cite{shh10}, our exact diagonalization result bears the signature of triangular vortex lattice \cite{tap13}.
We observe that with increasing angular momentum $L_{z}$, the system expands radially due to centrifugal force and consequently it becomes less dense.
The centrifugal effect is prominent for the multi-vortex states but becomes fairly small near the trap center. 
There is an upper limit on the rotational angular velocity in a harmonic trap. 
This limit is the trap frequency $\omega_{r}$ above which center-of-mass of the condensate destabilizes \cite{rps02}.
In the vicinity of high angular velocity $\Omega \lesssim \omega_{r}$, the centrifugal force influences the shape of the condensate.
Notice that the non-rotating $L_{z}=0$ state is more compact relative to the expanded higher angular momentum states.
\begin{figure}
\includegraphics[width=1.0\linewidth]{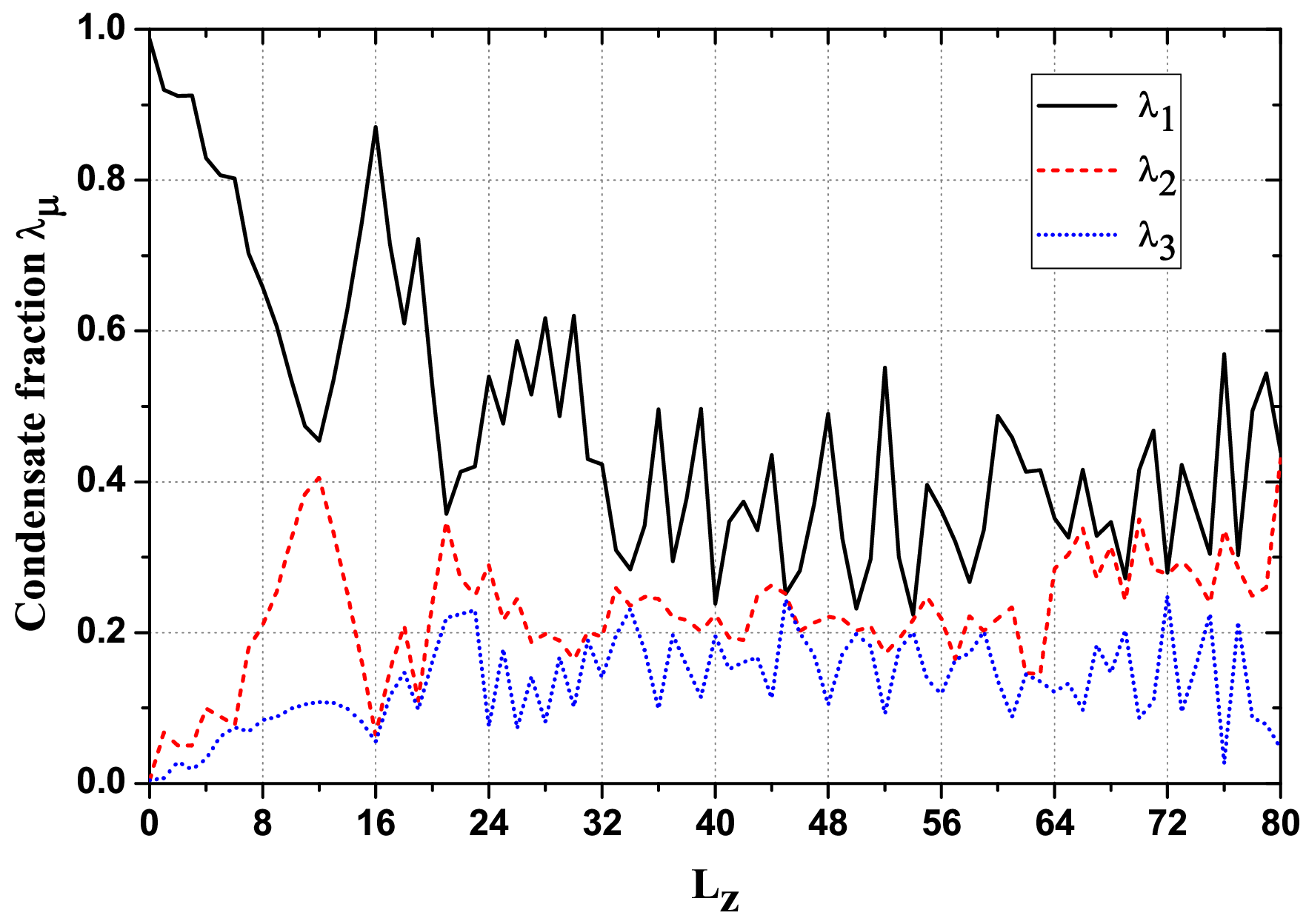} 
\caption{\label{fig:n16cond}(Color online) The largest three fractions corresponding to the eigenvalues $\lambda_{1}\ge\lambda_{2}\ge\lambda_{3}$ of the RSPDM in Eq.~(\ref{spd}) for $N=16$ bosons interacting via finite-range Gaussian potential, in the total angular momentum regime $0 \leq L_{z} \leq 5N$. The interaction parameters in Eq.~(\ref{gip}) has been chosen as $\mbox{g}_{2}=0.9151$ and $\sigma=0.1$. The macroscopic value of condensate fraction {\em i.e. }$\lambda_{1}$ has decreasing trend with $L_{z}$. In general, $\lambda_{1}$ is always larger than the eigenvalues of most eigenvectors within the ground states.}
\end{figure}
\\
\indent
In Fig.~\ref{fig:n16cond}, we have plotted three largest eigenvalues $\lambda_{1}>\lambda_{2}> \lambda_{3}$ of the RSPDM obtained from the ground state of the system {\it versus} the quantized total angular momentum in the regime $0 \leq L_{z} \leq 5N$.
It can be observed from the figure that nucleation of the first centered vortex $\left( \mbox{at } L_{z}=N\right)$ in the rotating Bose-condensate, does not occur through a smooth entrance of the vortex.
Indeed, the system passes through a correlated meta-stable angular momentum state at $L_{z}=12$, 
where the system achieves critically re-arrangement in configuration space, as evident by the von Neumann entropy ($S_{1}$) being maximum (see Fig.~\ref{fig:n16qent1}) and the two largest condensate fractions 
$\lambda_{1} (=0.4543)$, $\lambda_{2}(=0.4059)$ becoming comparable \cite{dbl09}.
Thus the system is preparing to undergo a critical phase transition from $L_{z}=11$ to $L_{z}=12$ state and the corresponding transition of single-particle quantum number $m_{1}=0$ to $m_{1}=1$ takes place, associated with largest condensate fraction $\lambda_{1}$.
A similar pattern is observed for the entry of second vortex at angular momentum state $L_{z}=21$ with $\lambda_{1} \sim \lambda_{2}$. 
The rotating system is again undergo a critical change in spatial correlation from $L_{z}=21$ to $L_{z}=22$ state and the corresponding transition of single-particle quantum number $m_{1}$ takes place that is from $m_{1}=1$ to $m_{1}=2$.
It is further observed that for a particular vortex state, the vorticity $m_{1}$ corresponding to largest condensate fraction $\lambda_{1}$ of RSPDM~(\ref{spd}) always attains the same value, irrespective of different values of quantum number $n$. 
For example, one may refer to Table~\ref{tab:n16_table2} where the single quantized vortex state has the same value of $m=1$ for quantum numbers $n=1$ and $3$.
We found that the angular momentum states with $L_{z}=$ 12, 22, 33, 43 \textit{etc.}, where the quantum jump of single-particle angular momentum $m_{1}$ takes place, correspond to the unstable states that provide a physical mechanism for the entry of vorticity into the condensate (refer to Table \ref{tab:cres}).
\begin{figure}[!htb]
\subfigure[$~L_{z}=0$]
{\includegraphics[width=0.32\linewidth]{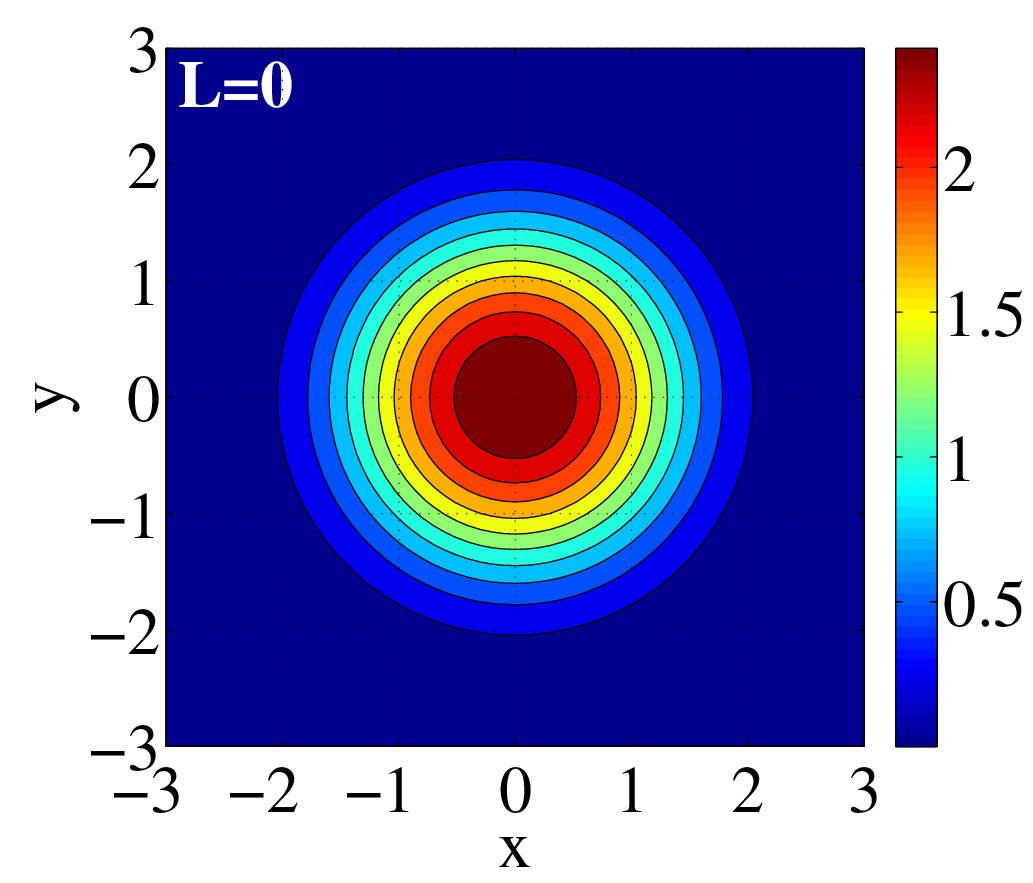}\label{fig:fv00}}
\subfigure[$~L_{z}=9$]
{\includegraphics[width=0.32\linewidth]{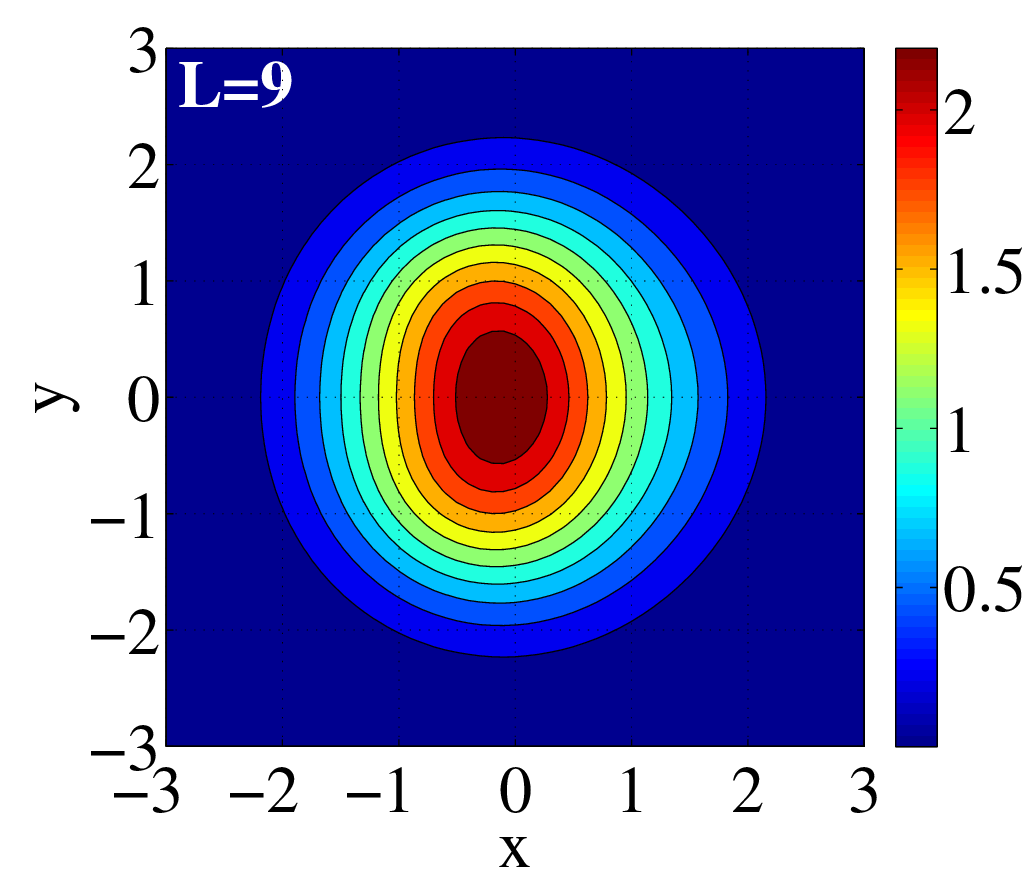}\label{fig:fv09}}
\subfigure[$~L_{z}=10$]
{\includegraphics[width=0.32\linewidth]{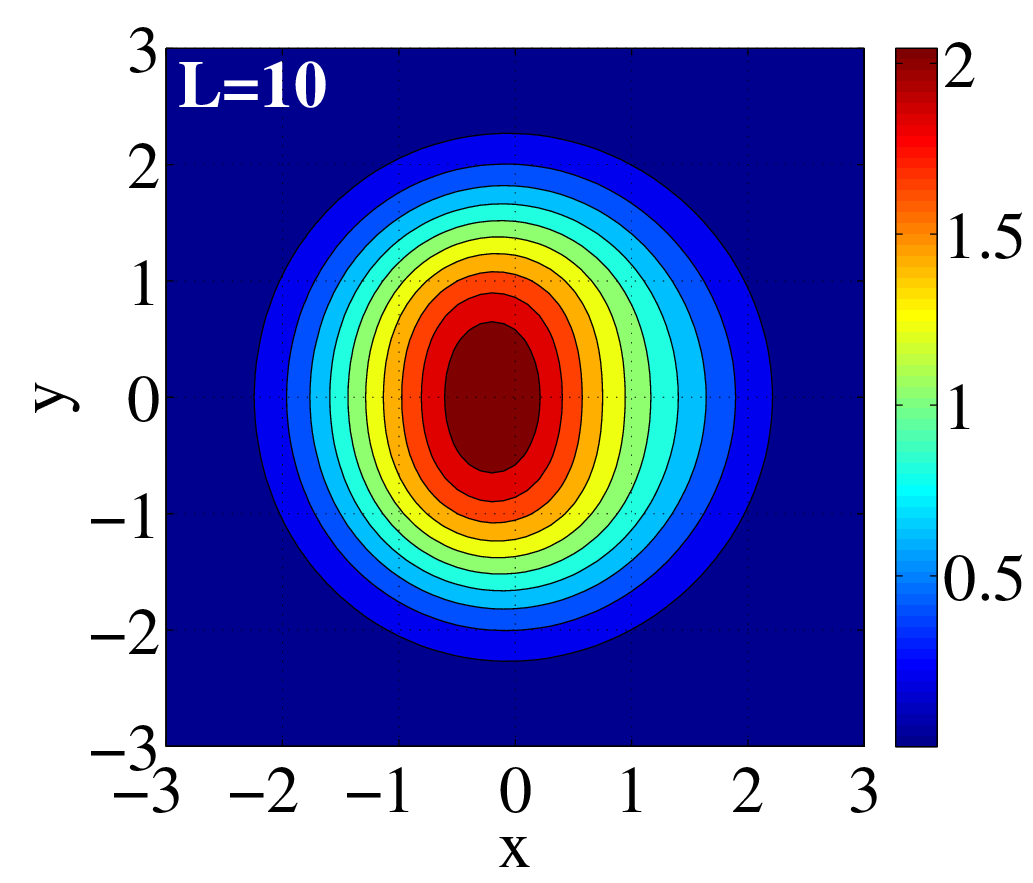}\label{fig:fv10}}
\subfigure[$~L_{z}=11$]
{\includegraphics[width=0.32\linewidth]{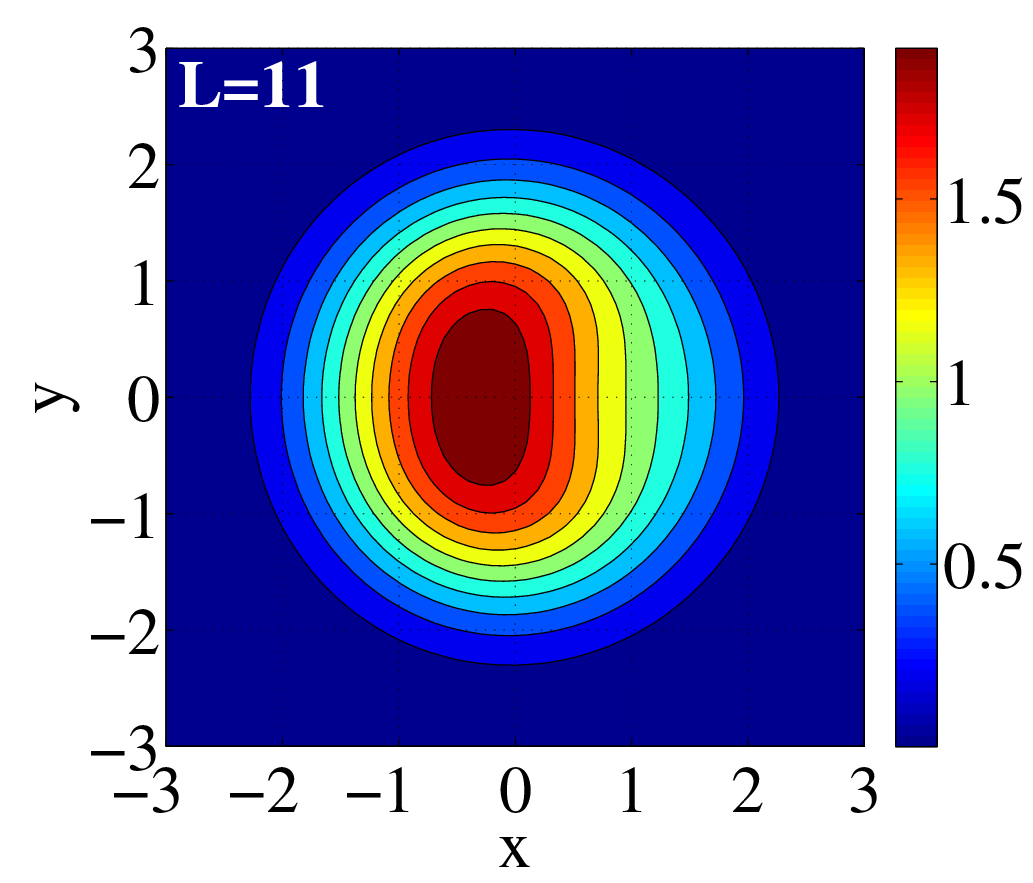}\label{fig:fv11}}
\subfigure[$~L_{z}=12$]
{\includegraphics[width=0.32\linewidth]{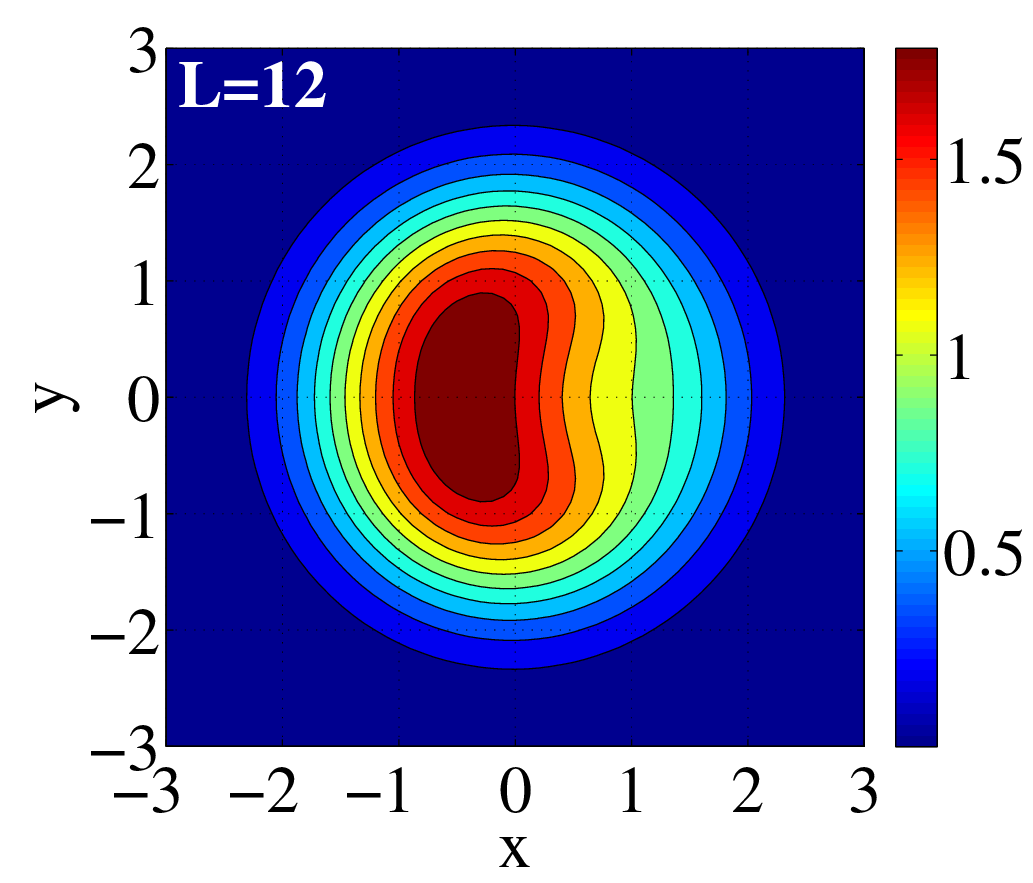}\label{fig:fv12}}
\subfigure[$~L_{z}=13$]
{\includegraphics[width=0.32\linewidth]{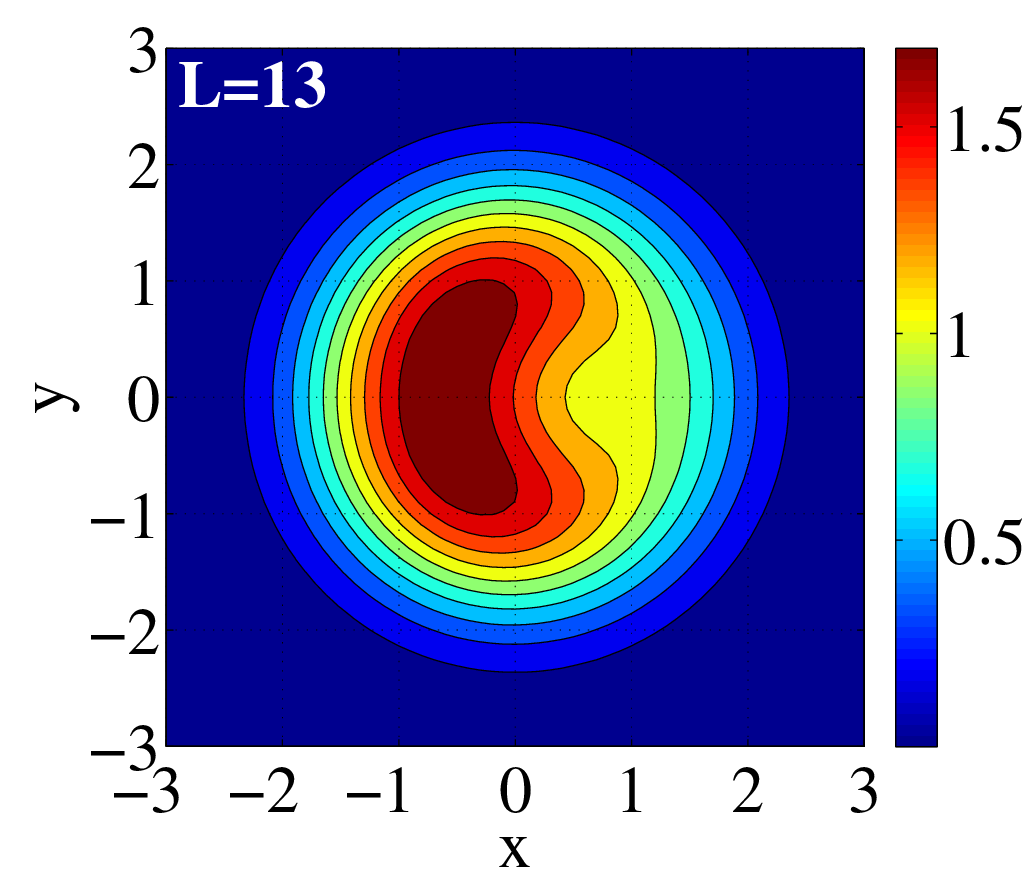}\label{fig:fv13}}
\subfigure[$~L_{z}=14$]
{\includegraphics[width=0.32\linewidth]{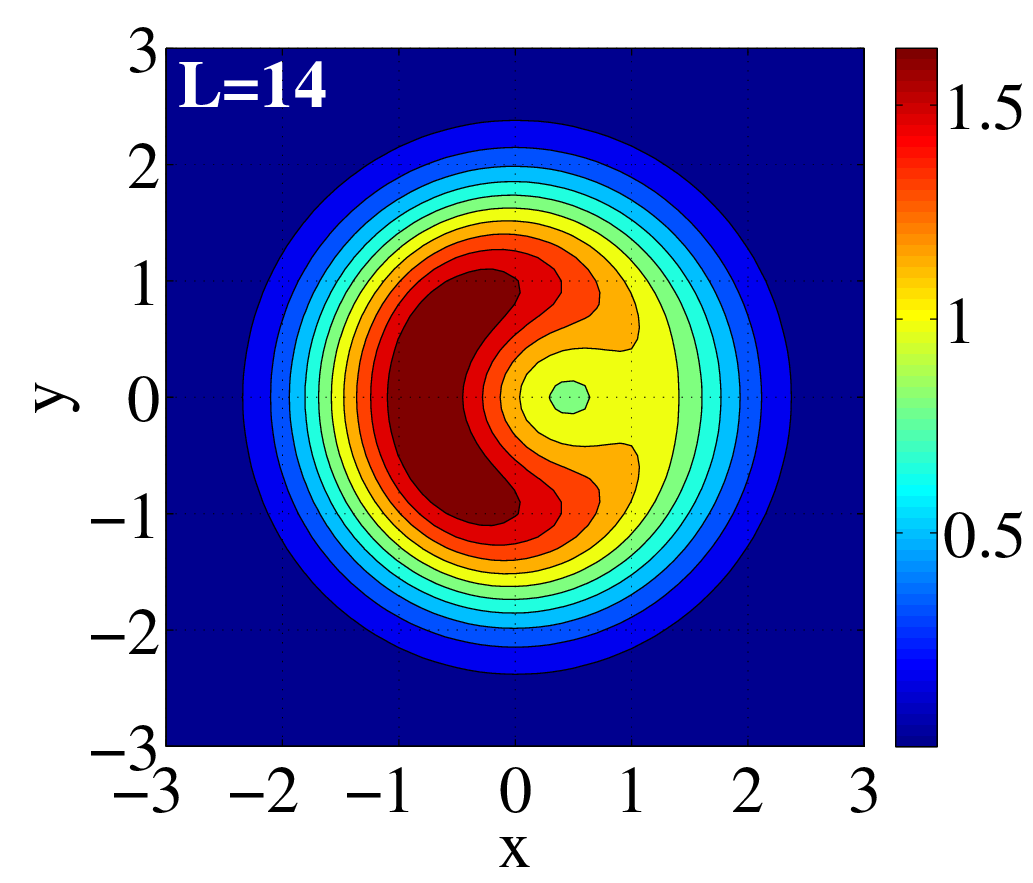}\label{fig:fv14}}
\subfigure[$~L_{z}=15$]
{\includegraphics[width=0.32\linewidth]{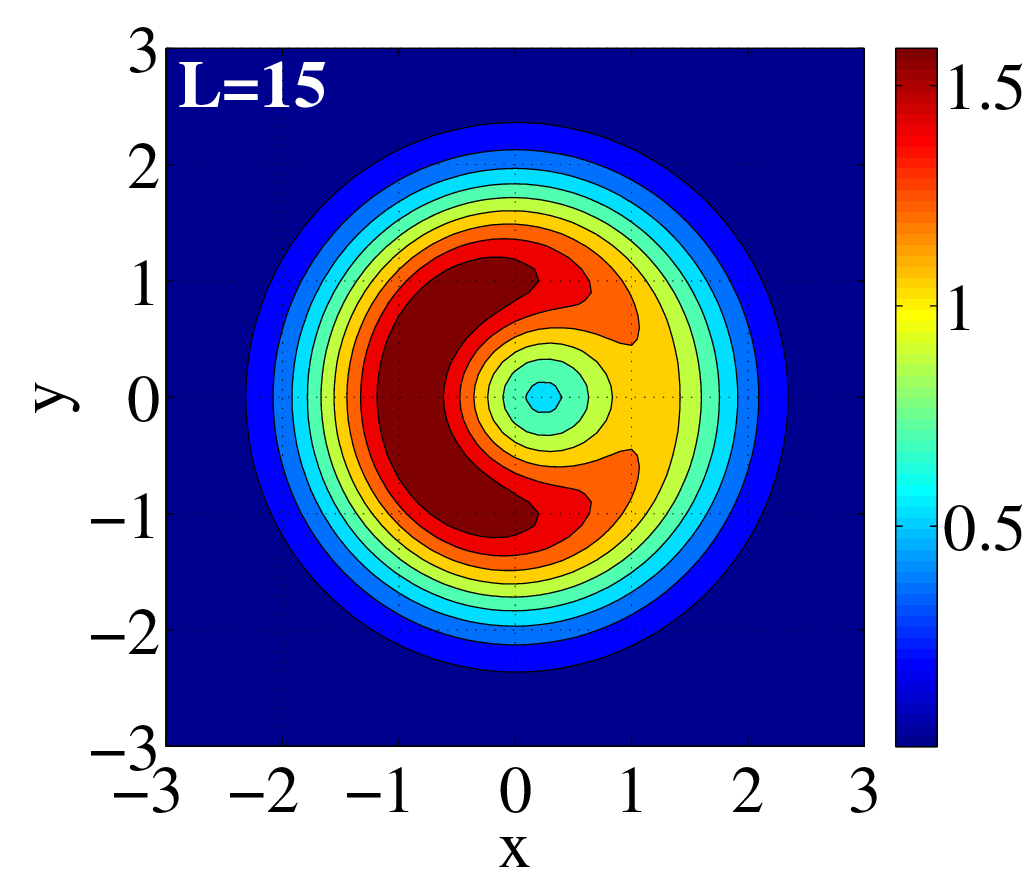}\label{fig:fv15}}
\subfigure[$~L_{z}=16$]
{\includegraphics[width=0.32\linewidth]{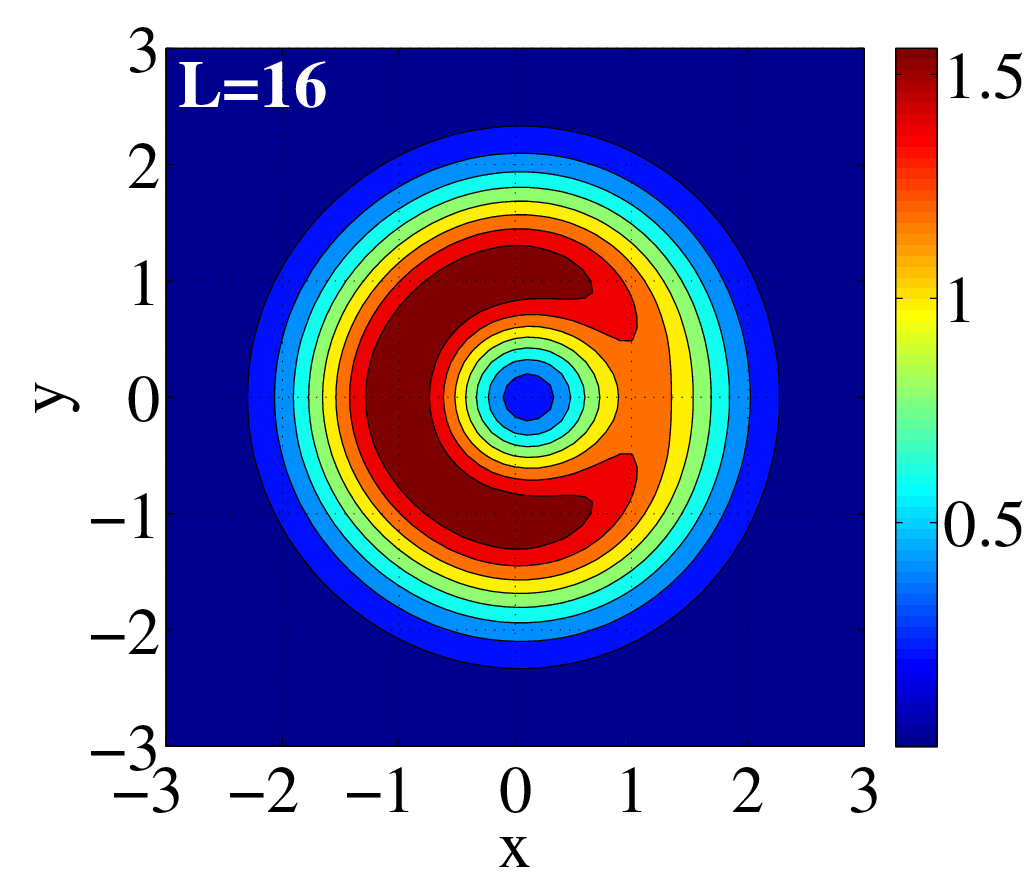}\label{fig:fv16}}
\caption{\label{fig:n16fv} (Color online) CPD contour plots for rotating Bose-condensed gas of $N=16$ particles in the subspace of total angular momentum regime $0 \le L_{z} \le N$ with interaction parameter $\mbox{g}_{2}=0.9151$ and range $\sigma=0.1$ of the Gaussian potential~(\ref{gip}). Each contour plot with ${\bf r}_{0}=(3,0)$, is an isosurface density profile viewed along the axis of rotation ($z$-axis).}
\end{figure}
\\
\indent
For better understanding, we focus on the angular momentum regime $0 \leq L_{z} \leq N$, in which the first vortex nucleates. 
The respective CPD plots calculated using Eq.(\ref{cpd}) are shown in Fig.~\ref{fig:n16fv}. 
The non-rotating state with angular momentum $L_{z}=0$ has Gaussian density profile along the axis of the trap as seen in Fig.~\ref{fig:fv00}. 
The density profile of this state has complete $x$-$y$ rotational symmetry of the harmonic trap.
The maximum number of bosons are confined around the trap center and occupy the non-rotating lowest single-particle state with angular momentum quantum number $m_{1}=0$ corresponding to largest condensate fraction $\lambda_{1}$ (the macroscopic eigenvalue of the RSPDM).
It may also be noted from Table~\ref{tab:n16_table1} that in the absence of external rotation the vortex-less state for $L_{z}=0$ is the most stable phase coherent state with the maximum value of degree of condensation ${C}_{d}=0.9857$ and minimum value of von Neumann entropy $S_{1}=0.0890$.
For an axially symmetric harmonic potential confining $N$ $(\gg 1)$ bosons, the CPD plots may be directly related to the position of the vortex with respect to the trap center. 
The regime $0 < L_{z} < N$, in above plots of Fig.~\ref{fig:n16fv}, describes an off-center unstable vortex state, for instance see Figs.~\ref{fig:fv09} to \ref{fig:fv15}.
The angular momentum state $L_{z}=N$ corresponds to a singly quantized vortex for which the axis of rotation coincides with the trap center as seen in the Fig.~\ref{fig:fv16}.
This state is identified by a high condensate fraction $\lambda_{1}$ (see Fig.~\ref{fig:n16cond}) with the single-particle quantum number $m_{1}=1$ as shown in Table~\ref{tab:n16_table1}, for which there is a peak in $C_{d}$ curve (Fig.~\ref{fig:n16doc}) and a corresponding dip in $S_{1}$ (Fig.~\ref{fig:n16qent1}).
The orientation of vortex can be represented as a measure of the possible spontaneous breaking of rotational symmetry.
In our system of harmonically confined $N=16$ rotating bosons, we observe that the circular symmetry is broken spontaneously, beyond $L_{z}=11$ as shown in Fig.~\ref{fig:fv11}.
The entry of first vortex is clearly seen to be from the periphery of the trap ($L_{z}=0$).
It is gradually moving inwards to its center ($L_{z}=N$) with increasing total angular momentum $L_{z}$.
\section{Summary and future work}\label{conc}
In summary, we have studied the novel phases of a rotating system of $N=16$ bosonic atoms confined in a quasi-2D harmonic trap and interacting repulsively via finite-range Gaussian potential.
Beyond the mean-field approximation, many-body ground states of the system are obtained using exact diagonalization method within subspaces of total angular momenta in the regime with $0 \leq L_{z} \leq 5N$.
By minimizing the energy of system in the co-rotating frame at zero-temperature, first, we have discussed the rotational critical angular velocity associated with the quantized total angular momentum, introduced as $L_{z}-\Omega$ stability curve.
From this we are able to provide information about the mechanical stability of angular momentum states, most notably the quantum phase transition between the stable vortex states.
In order to analyze the criterion for the existence of Bose-condensate in the rotating system considered here, we have calculated the first order correlation as embodied in reduced single-particle density matrix.
Many-body quantum correlation measured in terms of degree condensation and von Neumann entanglement entropy as a function of $L_{z}$, provide the direct evidence of quantum mechanically stable phase coherent states in consistent with the $L_{z}-\Omega$ stability curve.
Due to robust nature of quantum phase coherence in the rotating system, the stable vortex states corresponding to the successive critical angular velocities are found with lower value of entanglement entropy and higher value of degree of condensation.
Thereby internal structure of stable and unstable sates are  investigated by calculating the conditional probability distribution of angular momentum states in the co-rotating frame.
It is observed that the mechanism of formation and entrance of singly quantized vortices in the considered system is not smooth, indeed it passes through a critically re-arrangement in the configuration space, advocated by the analysis of internal structures in the unstable stable vortex states where the two largest condensate fractions become comparable.
The stable vortex states with upto seven singly quantized vortices are studied, possessing the $p$-fold discrete rotational symmetry with $p=2,3,4,5,6$.
The central vortex initially nucleated at $L_{z}=N=16$ state is peculiarly re-appeared at $L_{z}=71$ and $76$ states, forming a pentagon vortex pattern with vorticity $m_{1}=6$ and a hexagon vortex pattern with vorticity $m_{1}=7$, respectively.
In addition to the existence of five-fold rotational symmetry of vortex pattern in the confined system, our exact diagonalization results bear the signature of six-fold rotational symmetry precursor to the thermodynamically stable triangular vortex lattice. 
The complete exact diagonalization results for the model system considered here are summarised in a Table \ref{tab:cres}.
To make the physical picture of condensate density isosurface richer, we have also presented the CPD contour plots in the form of a movie \cite{video} depicting the formation of various vortex configurations with rotation from one stable state to the other.
As the results of the present study are significantly dependent on the interaction parameters chosen \cite{ahs01,ia20}, it would be interesting to investigate how these parameters, particularly the interaction range ($\sigma$), can be utilized as a probe for delving deeper into the internal structure and quantum mechanical stability of the vortex states \cite{ha23}.
Finally, we emphasize that the ground state properties, namely, critical angular velocity, degree of condensation and CPD, analyzed in the present work are the measurable quantities and hence experimentally relevant.
\appendix 
\section{Reduced single-particle density matrix}\label{sec:rspdm}
The zero-temperature reduced single-particle density matrix (RSPDM) is calculated from the variationally obtained exact $N$-body ground state wavefunction $\Psi\left({\bf r}_{1},{\bf r}_{2},\dots,{\bf r}_{N}\right)$, by integrating out the degree of freedoms of the $\left(N-1\right)$ particles, as:
\begin{eqnarray}
\rho \left({\bf r},{\bf r}^{\prime}\right) &=&\int \int \dots \int d{\bf r}_{2}\ d{\bf r}_{3}\dots d{\bf r}_{{N}} \nonumber\\
&&
\Psi^{\ast}({\bf r},{\bf r}_{2},{\bf r}_{3} \dots,{\bf r}_{{N}})\ \Psi({\bf r}^{\prime},{\bf r}_{2},{\bf r}_{3},\dots,{\bf r}_{{N}}) \nonumber \\
&\equiv & \sum_{{\bf n},{\bf n}^{\prime}}\ \rho_{_{{\bf n},{\bf n}^{\prime}}}\ u^{\ast}_{\bf n}\left({\bf r}\right)u_{{\bf n}^{\prime }}\left({\bf r}^{\prime }\right).  
\end{eqnarray}
The above expression is written in terms of single-particle basis functions $u_{\bf n}\left({\bf r}\right)$ with quantum number ${\bf n}\equiv \left(n,m\right)$.
Being hermitian, this can be diagonalized to give 
\begin{equation}
\rho \left({\bf r},{\bf r}^{\prime}\right) = \sum_{\mu }\lambda_{\mu } \ \chi^{\ast}_{\mu }\left({\bf r}\right)
\chi_{\mu } \left({\bf r}^{\prime }\right),
\label{spd}
\end{equation}  
where $
\chi_{\mu }\left({\bf r}\right) \equiv \sum_{\bf n} c^{\mu }_{\bf n} \, {u}_{\bf n}\left({\bf r}\right)$ and $\sum_{\mu }\lambda_{\mu }=1$
with $ 1\geq\lambda_{1}\geq\lambda_{2}\geq\cdots \geq\lambda_{\mu }\geq\cdots \geq 0 $.
The $\left\{\lambda_{\mu }\right\}$ are the eigenvalues, ordered as above, and $\left\{ \chi_{\mu }\left({\bf r}\right)\right\}$ are the corresponding eigenvectors of the RSPDM (\ref{spd}); each $\mu$ defines a fraction of the BEC.
According to the Penrose-Onsager criterion, ``B.E. condensation is said to be present whenever the largest eigenvalue of the reduced single-particle density matrix is an extensive rather than an intensive quantity" \cite{po56}.
Conveniently, a fragmented state is defined as a state for which there exits more than one macroscopically large eigenvalue of the RSPDM \cite{nj82}. 
The fragmented ground state can as well be viewed as a quantum mechanical average over the symmetry broken single condensed states \cite{mhu06}.
\begin{table}[!htb]
\caption{\label{tab:n16_table2} The largest macroscopic eigenvalue $\lambda_{1}>\lambda_{(\mu \geq 2)}$ in Eq.~(\ref{spd}) and the weight $\left|c^{1}_{n,m_{1}}\right|^{2}$ of the corresponding eigenvectors with their quantum numbers $\left(n,m_{1}\right)$ are reported at each respective critical angular velocity $\Omega_{{\bf c}}\left(L_{z}^{i}\right),~i=1,2,3,\dots$ for $N=16$ bosons, in given subspaces of the total angular momentum $L_{z}^{i}$. The interaction strength $\mbox{g}_{2}=0.9151$ and range $\sigma=0.1$ of the Gaussian potential~(\ref{gip}).}
\begin{ruledtabular}
\begin{tabular}{cccccc}
$L_{z}^{i}$ & $\Omega_{\bf c}\left(L_{z}^{i}\right)$ & ($n,m_{1}$) & ${\left|c^{1}_{n,m_{1}}\right|}^{2}$ & $\lambda_{1}$ \\ 
\hline 
\multirow{2}{*}{0} & \multirow{2}{*}{0.0000} & 0, 0 & 0.9467 & \multirow{2}{*}{0.9872}\\ 
 & & 2, 0 & 0.0532 & \\ \hline
\multirow{2}{*}{16} & \multirow{2}{*}{0.7745} & 1, 1 & 0.9890 & \multirow{2}{*}{0.8707} \\
 & & 3, 1 & 0.0109 & \\ \hline
\multirow{2}{*}{26} & \multirow{2}{*}{0.9028} & 2, 2 & 0.9898 & \multirow{2}{*}{0.5869} \\
 & & 4, 2 & 0.0101 & \\ \hline
 \multirow{2}{*}{28} & \multirow{2}{*}{0.9048} & 2, 2 & 0.9922 & \multirow{2}{*}{0.6171} \\
 & & 4, 2 & 0.0078 & \\ \hline 
 \multirow{2}{*}{32} & \multirow{2}{*}{0.9218} & 2, 2 & 0.9953 & \multirow{2}{*}{0.4231} \\
 & & 4, 2 & 0.0046 & \\ \hline 
 \multirow{2}{*}{35} & \multirow{2}{*}{0.9236} & 3, 3 & 0.9946 & \multirow{2}{*}{0.3420} \\
 & & 5, 3 & 0.0053 & \\ \hline 
 \multirow{2}{*}{36} & \multirow{2}{*}{0.9273} & 3, 3 & 0.9948 & \multirow{2}{*}{0.4964} \\
 & & 5, 3 & 0.0051 & \\ \hline
 \multirow{2}{*}{48} & \multirow{2}{*}{0.9423} & 4, 4 & 0.9975 & \multirow{2}{*}{0.4901} \\
 & & 6, 4 & 0.0024 & \\ \hline
 \multirow{2}{*}{52} & \multirow{2}{*}{0.9740} & 4, 4 & 0.9979 & \multirow{2}{*}{0.5513} \\
 & & 6, 4 & 0.0020 & \\ \hline
 60 & 0.9805 & 5, 5 & 1.0000 & 0.4873 \\ \hline
 71 & 0.9835 & 6, 6 & 1.0000 & 0.4681 \\ \hline
 76 & 0.9857 & 7, 7 & 1.0000 & 0.5691  \\
\end{tabular}
\end{ruledtabular} 
\end{table}
\\
\indent
For the quasi-2D system, where $z$ degree of freedom in the RSPDM has been traced out, Eq.~(\ref{spd}) reduces to 
\begin{equation}
\rho\left(r,\theta;r^{\prime},\theta^{\prime}\right)=\sum_{\mu} \lambda_{\mu} \chi_{\mu}\left(r,\theta\right) \chi_{\mu}^{\ast}\left(r^{\prime},{\theta}^{\prime}\right),
\label{spd2}
\end{equation}
where index $\mu=1,2,3,\dots$ are the new labels for the single-particle basis quantum numbers corresponding to each fragment of the condensate.
We further have found from our calculation that the eigenvector corresponding to a given fragment has the form 
\begin{equation*}
\chi_{\mu}\left(r,\theta\right)=\left(\sum_{n} c^{\mu}_{n,m_{\mu}}~ f^{m_{\mu}}_{n}\left(r\right)\right)e^{im_{\mu}\theta} 
\equiv f_{m_{\mu}}(r)e^{im_{\mu}\theta}
\label{spv}
\end{equation*}
where
\begin{equation*}
f^{m_{\mu}}_{n}(r)=\sqrt{\frac{\left(\frac{1}{2}\left\{ n-|m_{\mu}| \right\} \right)!}{\pi\left(\frac{1}{2}\left\{ n+|m_{\mu}| \right\} \right)!}}\ e^{-\frac{1}{2}r^{2}} r^{|m_{\mu}|} L^{|m_{\mu}|}_{\frac{1}{2}(n-|m_{\mu}|)}(r^{2})\nonumber
\end{equation*}
For a particular total angular momentum $L_{z}$-state, every fraction $\lambda_{\mu}$ of the RSPDM is characterized by a value of single-particle angular momentum quantum number $m_{\mu}$.
It is observed that $m_{\mu}$ corresponding to a condensate fraction always takes the same value, irrespective of quantum number $n$.
For example, one may refer to Table~\ref{tab:n16_table2} where the single quantized vortex state have the same value of $m_{\mu}=1$ for quantum numbers $n=1$ and $3$ corresponding to $\mu=1$.
\\
\indent
For a many-body system, the realization of BEC corresponds to a single eigenvalue $\lambda_{1}$ becoming significantly larger than the remaining eigenvalues, {\em i.e.} $\lambda_{1} \gg \lambda_{(\mu \geq 2)}\ \forall \ \mu$, and identified as the largest condensate fraction.
The corresponding eigenvector $\chi_{1}\left({\bf r}\right)=\sum_{\bf n}c^{1}_{\bf n} u_{\bf n}\left({\bf r}\right)$, takes the role of the macroscopic order parameter of the system, in the mean-field Gross-Pitaevskii description. 
The quantum number $m_{1}$, corresponding to the largest fractional amplitude $\left|c^{1}_{\bf n}\right|^{2}$ of single-particle function $u_{\bf n}\left({\bf r}\right)$, is recognized as vorticity of the condensate.
\section{Degree of condensation}\label{sec:doc}
It is to be noted that the usual definition of condensation for a macroscopic system, given by the largest eigenvalue $\lambda_{1}$ of the RSPDM~(\ref{spd}), is not appropriate for systems with small number of particles being studied here \cite{dbo07,iasl15,ia20}.
For example, in the absence of condensation, there is no macroscopic occupation of a single quantum state and all levels are equally occupied, such a definition would imply a condensate though with small magnitude. 
To circumvent this situation, one introduces a quantity which is sensitive to the loss of macroscopic occupation called the degree of condensation defined as
\begin{equation}
C_{d} =\lambda_{1}-\bar{\lambda}
\label{doc}
\end{equation}
where $\bar{\lambda}=\frac{1}{q-1}\sum_{\mu=2}^{q} \lambda_{\mu}$ is the mean of the rest of eigenvalues. 
It can be seen that the degree of condensation, defined as in Eq.~(\ref{doc}), approaches zero for equal eigenvalues, as one would expect.
In Fig.~\ref{fig:n16doc}, we present the variation of degree of condensation $C_{d}$ with total angular momentum $L_{z}$ for $N=16$ bosons.
\begin{table*}[!]
\caption{\label{tab:cres}\textbf{Complete exact diagonalization results:}
For $N=16$ bosons in given subspaces of total angular momentum $0 \le L_{z} \le 5N$, the lowest eigenenergy $E^{lab}(L_{z})$ in units of $\hbar \omega_{r}$ of the states in the laboratory frame, the values of critical angular velocity $\Omega_{{\bf c}}\left(L_{z}^{i}\right)$ with $p$-fold rotational symmetry of stable vortex states, the largest three eigenvalues $\lambda_{1}>\lambda_{2}>\lambda_{3}$ and the corresponding single-particle quantum numbers $\left(n_{1},m_{1}\right),\left(n_{2},m_{2}\right)$ and $\left(n_{3},m_{3}\right)$ in the RSPDM~(\ref{spd}) for the ground state of the rotating BEC. Also given are the values of degree of condensation ${C}_{d}$~(\ref{doc}) and von Neumann entropy $S_{1}$~(\ref{vent}) of the many-body quantum states. Here, interaction parameter $\mbox{g}_{2}=0.9151$ and range $\sigma=0.1$ of the repulsive Gaussian interaction potential~(\ref{gip}). See supplementary material \cite{video} for the internal structure of each tabulated angular momentum $L_{z}$-state (includes the stable and unstable states) which clearly shows the nucleation and entry of the vortices alongwith its arrangement to the various stable vortex configurations in a harmonically confined system.}
\begin{ruledtabular}
\begin{tabular}{cccccccccccc}
$L_{z}$ & $E^{lab}(L_{z})$ & $\Omega_{\bf c}$ & $p$ & $(n_{1},m_{1})$ & $\lambda_{1}$ & $(n_{2},m_{2})$ & $\lambda_{2}$ & $(n_{3},m_{3})$ & $\lambda_{3}$ & $C_{d}$ & $S_{1}$ \\ 
\hline 
0 & 47.09788 & 0.0 & - & 0, 0 & 0.9872 & 1, 1 & 0.0043 & 1, -1 & 0.0043 & 0.9857 & 0.0890 \\
1 & 48.10784 &&& 0, 0 & 0.9197 & 1, 1 & 0.0677 & 1, -1 & 0.0072 & 0.9108 & 0.3313 \\
2 & 48.90396 &&& 0, 0 & 0.9119 & 2, 2 & 0.0505 & 1, 1 & 0.0287 & 0.9022 & 0.3915 \\
3 & 49.59215 &&& 0, 0 & 0.9124 & 3, 3 & 0.0502 & 1, 1 & 0.0186 & 0.9026 & 0.4047 \\
4 & 50.52231 &&& 0, 0 & 0.8292 & 1, 1 & 0.0998 & 3, 3 & 0.0326 & 0.8102 & 0.6535 \\
5 & 51.30136 &&& 0, 0 & 0.8068 & 1, 1 & 0.0889 & 2, 2 & 0.0622 & 0.7853 & 0.7237 \\
6 & 52.04369 &&& 0, 0 & 0.8023 & 1, 1 & 0.0774 & 3, 3 & 0.0743 & 0.7803 & 0.7370 \\
7 & 52.86634 &&& 0, 0 & 0.7029 & 1, 1 & 0.1794 & 2, 2 & 0.0688 & 0.6699 & 0.9166 \\
8 & 53.62644 &&& 0, 0 & 0.6575 & 1, 1 & 0.2101 & 2, 2 & 0.0840 & 0.6194 & 0.9849 \\
9 & 54.37954 &&& 0, 0 & 0.6061 & 1, 1 & 0.2546 & 2, 2 & 0.0883 & 0.5624 & 1.0435 \\      
10 & 55.13152 &&& 0, 0 & 0.5364 & 1, 1 & 0.3235 & 2, 2 & 0.0991& 0.4848 & 1.0825 \\
11 & 55.87272 &&& 0, 0 & 0.4738 & 1, 1 & 0.3839 & 2, 2 & 0.1052 & 0.4153 & 1.1011 \\
12 & 56.60990 &&& 1, 1 & 0.4543 & 0, 0 & 0.4059 & 2, 2 & 0.1078 & 0.3936 & 1.0931 \\
13 & 57.34398 &&& 1, 1 & 0.5351 & 0, 0 & 0.3323 & 2, 2 & 0.1065 & 0.4834 & 1.0504 \\
14 & 58.07544 &&& 1, 1 & 0.6293 & 0, 0 & 0.2518 & 2, 2 & 0.0988 & 0.5880 & 0.9591 \\
15 & 58.80387 &&& 1, 1 & 0.7432 & 0, 0 & 0.1612 & 2, 2 & 0.0818 & 0.7147 & 0.7888 \\
16 & 59.49139 & 0.7746  & - & 1, 1 & 0.8707 & 0, 0 & 0.0643 & 2, 2 & 0.0554 & 0.8589 & 0.5128 \\
17 & 60.49352 &&& 1, 1 & 0.7152 & 2, 2 & 0.1507 & 0, 0 & 0.1157 & 0.6893 & 0.8673 \\
18 & 61.47371 &&& 1, 1 & 0.6099 & 2, 2 & 0.2091 & 0, 0 & 0.1475 & 0.5745 & 1.0624 \\
19 & 62.29758 &&& 1, 1 & 0.7218 & 2, 2 & 0.1094 & 0, 0 & 0.0981 & 0.6966 & 0.9521 \\
20 & 63.24292 &&& 1, 1 & 0.5259 & 2, 2 & 0.2405 & 0, 0 & 0.1625 & 0.4828 & 1.2281 \\
21 & 64.16139 &&& 1, 1 & 0.3575 & 2, 2 & 0.3470 & 0, 0 & 0.2193 & 0.2990 & 1.3323 \\
22 & 65.01723 &&& 2, 2 & 0.4133 & 0, 0 & 0.2721 & 1, 1 & 0.2248 & 0.3599 & 1.3392 \\
23 & 65.92312 &&& 2, 2 & 0.4202 & 1, 1 & 0.2493 & 0, 0 & 0.2295 & 0.3675 & 1.3658 \\
24 & 66.75711 &&& 2, 2 & 0.5394 & 0, 0 & 0.2898 & 1, 1 & 0.0758 & 0.4975 & 1.1796 \\
25 & 67.70235 &&& 2, 2 & 0.4773 & 0, 0 & 0.2167 & 1, 1 & 0.1783 & 0.4298 & 1.3584 \\
26 & 68.52023 & 0.9028 & 2 & 2, 2 & 0.5869 & 0, 0 & 0.2446 & 4, 4 & 0.0730 & 0.5493 & 1.1478 \\
27 & 69.51889 &&& 2, 2 & 0.5156 & 0, 0 & 0.1868 & 1, 1 & 0.1420 & 0.4716 & 1.3388 \\
28 & 70.32997 & 0.9048 & 2 & 2, 2 & 0.6171 & 0, 0 & 0.1983 & 4, 4 & 0.0810 & 0.5823 & 1.1402 \\
29 & 71.37778 &&& 2, 2 & 0.4870 & 0, 0 & 0.1897 & 3, 3 & 0.1673 & 0.4404 & 1.3717 \\
30 & 72.19763 &&& 2, 2 & 0.6202 & 0, 0 & 0.1641 & 4, 4 & 0.1019 & 0.5857 & 1.1616 \\
31 & 73.24433 &&& 2, 2 & 0.4303 & 3, 3 & 0.2003 & 0, 0 & 0.1903 & 0.3785 & 1.4432 \\
32 & 74.01746 & 0.9218 & 2 & 2, 2 & 0.4238 & 0, 0 & 0.1945 & 4, 4 & 0.1409 & 0.3707 & 1.5611 \\
33 & 74.94875 &&& 3, 3 & 0.3093 & 0, 0 & 0.2592 & 2, 2 & 0.1961 & 0.2465 & 1.6453 \\
34 & 75.87474 &&& 3, 3 & 0.2836 & 0, 0 & 0.2358 & 2, 2 & 0.2314 & 0.2185 & 1.6561 \\
35 & 76.78844 & 0.9236 & 3 & 3, 3 & 0.3420 & 0, 0 & 0.2474 & 2, 2 & 0.1764 & 0.2821 & 1.6116 \\
36 & 77.71583 & 0.9274 & 3 & 3, 3 & 0.4964 & 0, 0 & 0.2446 & 2, 2 & 0.0992 & 0.4506 & 1.3889 \\
37 & 78.69205 &&& 3, 3 & 0.2944 & 0, 0 & 0.2212 & 2, 2 & 0.1978 & 0.2303 & 1.6952 \\
38 & 79.61633 &&& 3, 3 & 0.3789 & 0, 0 & 0.2165 & 2, 2 & 0.1555 & 0.3224 & 1.6003 \\
39 & 80.54489 &&& 3, 3 & 0.4969 & 0, 0 & 0.2013 & 4, 4 & 0.1143 & 0.4512 & 1.4305 \\
40 & 81.53841 &&& 0, 0 & 0.2385 & 3, 3 & 0.2242 & 4, 4 & 0.1957 & 0.1693 & 1.7301 \\
\end{tabular}
\end{ruledtabular}
\end{table*}
\begin{table*}[!]
\begin{ruledtabular}
\begin{tabular}{cccccccccccc}
$L_{z}$ & $E^{lab}(L_{z})$ & $\Omega_{\bf c}$ & $p$ & $(n_{1},m_{1})$ & $\lambda_{1}$ & $(n_{2},m_{2})$ & $\lambda_{2}$ & $(n_{3},m_{3})$ & $\lambda_{3}$ & $C_{d}$ & $S_{1}$ \\ 
\hline 
41 & 82.50266 &&& 3, 3 & 0.3470 & 0, 0 & 0.1931 & 4, 4 & 0.1514 & 0.2876 & 1.6519 \\
42 & 83.43884 &&& 3, 3 & 0.3737 & 0, 0 & 0.1903 & 4, 4 & 0.1610 & 0.3168 & 1.6248 \\
43 & 84.39746 &&& 4, 4 & 0.3359 & 0, 0 & 0.2485 & 3, 3 & 0.1668 & 0.2755 & 1.6178 \\
44 & 85.33349 &&& 4, 4 & 0.4356 & 0, 0 & 0.2628 & 5, 5 & 0.1130 & 0.3843 & 1.4697 \\
45 & 86.34788 &&& 0, 0 & 0.2516 & 4, 4 & 0.2513 & 5, 5 & 0.2437 & 0.1835 & 1.6526 \\
46 & 87.32416 &&& 4, 4 & 0.2822 & 0, 0 & 0.2020 & 5, 5 & 0.1993 & 0.2170 & 1.6884 \\
47 & 88.24472 &&& 4, 4 & 0.3709 & 0, 0 & 0.2134 & 5, 5 & 0.1699 & 0.3138 & 1.5786 \\
48 & 89.02375 & 0.9423 & 4 & 4, 4 & 0.4901 & 0, 0 & 0.2212 & 3, 3 & 0.1047 & 0.4437 & 1.4733 \\
49 & 90.05830 &&& 4, 4 & 0.3245 & 0, 0 & 0.2184 & 5, 5 & 0.1716 & 0.2630 & 1.6975 \\
50 & 91.04188 &&& 4, 4 & 0.2318 & 0, 0 & 0.2024 & 3, 3 & 0.1981 & 0.1619 & 1.8293 \\
51 & 91.99455 &&& 4, 4 & 0.2975 & 3, 3 & 0.2082 & 0, 0 & 0.1819 & 0.2336 & 1.7489 \\
52 & 92.92004 & 0.9741 & 4 & 4, 4 & 0.5513 & 0, 0 & 0.1720 & 3, 3 & 0.0924 & 0.5105 & 1.3965 \\
53 & 93.94809 &&& 4, 4 & 0.3003 & 5, 5 & 0.1918 & 0, 0 & 0.1772 & 0.2366 & 1.7494 \\   
54 & 94.93058 &&& 5, 5 & 0.2239 & 0, 0 & 0.2165 & 4, 4 & 0.2007 & 0.1533 & 1.7953 \\
55 & 95.89720 &&& 5, 5 & 0.3958 & 0, 0 & 0.2475 & 6, 6 & 0.1387 & 0.3408 & 1.6103 \\
56 & 96.87361 &&& 6, 6 & 0.3616 & 1, 1 & 0.2189 & 0, 0 & 0.1187 & 0.3035 & 1.7292 \\
57 & 97.88129 &&& 6, 6 & 0.3205 & 5, 5 & 0.1644 & 0, 0 & 0.1644 & 0.2587 & 1.8109 \\
58 & 98.86454 &&& 6, 6 & 0.2669 & 5, 5 & 0.2219 & 0, 0 & 0.1729 & 0.2003 & 1.8200 \\
59 & 99.81495 &&& 5, 5 & 0.3365 & 6, 6 & 0.2018 & 0, 0 & 0.2016 & 0.2762 & 1.6892 \\
60 & 100.76424 & 0.9805 & 5 & 5, 5 & 0.4873 & 0, 0 & 0.2191 & 6, 6 & 0.1366 & 0.4407 & 1.4345 \\
61 & 101.75677 &&& 6, 6 & 0.4588 & 1, 1 & 0.2343 & 0, 0 & 0.0881 & 0.4095 & 1.5379 \\
62 & 102.77963 &&& 6, 6 & 0.4135 & 0, 0 & 0.1461 & 5, 5 & 0.1447 & 0.3602 & 1.6910 \\
63 & 103.79201 &&& 6, 6 & 0.4154 & 5, 5 & 0.1453 & 0, 0 & 0.1354 & 0.3623 & 1.6936 \\
64 & 104.73899 &&& 1, 1 & 0.3513 & 7, 7 & 0.2848 & 4, 4 & 0.1211 & 0.2924 & 1.6295 \\
65 & 105.71752 &&& 1, 1 & 0.3261 & 6, 6 & 0.3038 & 7, 7 & 0.1324 & 0.2648 & 1.6257 \\
66 & 106.66694 &&& 6, 6 & 0.4160 & 1, 1 & 0.3380 & 7, 7 & 0.0976 & 0.3629 & 1.4282 \\
67 & 107.67820 &&& 1, 1 & 0.3281 & 7, 7 & 0.2722 & 6, 6 & 0.1835 & 0.2670 & 1.6344 \\
68 & 108.70366 &&& 1, 1 & 0.3464 & 7, 7 & 0.3149 & 6, 6 & 0.1468 & 0.2870 & 1.5511 \\
69 & 109.66932 &&& 1, 1 & 0.2719 & 7, 7 & 0.2426 & 6, 6 & 0.2034 & 0.2057 & 1.7417 \\
70 & 110.60461 &&& 7, 7 & 0.4159 & 1, 1 & 0.3503 & 6, 6 & 0.0862 & 0.3628 & 1.4117 \\
71 & 111.58246 & 0.9835 & 5 & 6, 6 & 0.4681 & 1, 1 & 0.2846 & 7, 7 & 0.1092 & 0.4197 & 1.3919 \\
72 & 112.60024 &&& 7, 7 & 0.2796 & 1, 1 & 0.2773 & 6, 6 & 0.2488 & 0.2142 & 1.6297 \\
73 & 113.59806 &&& 7, 7 & 0.4227 & 1, 1 & 0.2954 & 6, 6 & 0.0947 & 0.3702 & 1.4892 \\
74 & 114.62161 &&& 7, 7 & 0.3633 & 1, 1 & 0.2748 & 6, 6 & 0.1545 & 0.3054 & 1.5792 \\
75 & 115.58554 &&& 6, 6 & 0.3044 & 7, 7 & 0.2394 & 1, 1 & 0.2248 & 0.2412 & 1.6594 \\
76 & 116.51123 & 0.9857 & 6 & 7, 7 & 0.5691 & 1, 1 & 0.3356 & 6, 6 & 0.0265 & 0.5299 & 1.0564 \\
77 & 117.56864 &&& 6, 6 & 0.3029 & 7, 7 & 0.2853 & 1, 1 & 0.2142 & 0.2395 & 1.6227 \\
78 & 118.56103 &&& 7, 7 & 0.4941 & 1, 1 & 0.2486 & 6, 6 & 0.0874 & 0.4481 & 1.4388 \\
79 & 119.55921 &&& 7, 7 & 0.5436 & 1, 1 & 0.2600 & 4, 4 & 0.0786 & 0.5021 & 1.2854 \\
80 & 120.55276 &&& 2, 2 & 0.4397 & 8, 8 & 0.4307 & 5, 5 & 0.0466 & 0.3888 & 1.1859 \\
\end{tabular}
\end{ruledtabular}
\end{table*}
\acknowledgments
\noindent
We would like to thank late Professor M. Rafat (JMI) for helpful discussions.
%

\end{document}